\documentclass[11pt]{article}  

\usepackage{graphicx}
\usepackage{amsmath}
\usepackage{subcaption}
\usepackage[a4paper, margin=2cm]{geometry}
\usepackage[colorlinks=true,     linkcolor=blue, citecolor=blue, filecolor=blue, urlcolor=blue]{hyperref}
\usepackage{txfonts}

\usepackage[nonumberlist,sanitizesort,nogroupskip]{glossaries}
\renewcommand{\glossarysection}[2][]{}
\makenoidxglossaries
\newglossaryentry{alpha_corr}{
	name=\ensuremath{\alpha_{\text{corr}}},
	description={Constant linking the response of the correlator to correlated inputs ($f_{\text{corr}}$)}
}

\newglossaryentry{alpha_IR}{
	name=\ensuremath{\alpha_{\text{IR}}},
	description={Constant for the heterodyne infrared detection linking the optical signals to down-converted signal}
}

\newglossaryentry{A}{
	name=\ensuremath{A},
	description={Constant for the response of the correlator to uncorrelated inputs ($f_{\text{noise}}$)}
}

\newglossaryentry{B_vectot}{
	name=\ensuremath{\vec{B}},
	description={Interferometric baseline}
}

\newglossaryentry{B}{
	name=\ensuremath{B},
	description={Constant for the response of the correlator to uncorrelated inputs ($f_{\text{noise}}$)}
}

\newglossaryentry{C}{
	name=\ensuremath{C},
	description={Constant for the response of the correlator to uncorrelated inputs ($f_{\text{noise}}$)}
}

\newglossaryentry{beta_k}{
	name=\ensuremath{\beta_k},
	description={Constant for Mach-Zender Modulator \ensuremath{k} (\ensuremath{\beta_k=\pi/V_{\pi,k}})}
}

\newglossaryentry{delta_nu_det}{
	name=\ensuremath{\Delta\nu_{\text{det}}},
	description={Detector bandwidth}
}

\newglossaryentry{delta_nu_noise}{
	name=\ensuremath{\Delta\nu_{\text{noise}}},
	description={Bandwidth of noises $w_1$ and $w_2$}
}

\newglossaryentry{delta_omega_k}{
	name=\ensuremath{\Delta\omega_{k}},
	description={Frequency of Acousto-Optic Frequency Shifter \ensuremath{k}}
}

\newglossaryentry{delta_Phi}{
	name=\ensuremath{\Delta\Phi},
	description={Total accumulated phase difference between $s_1$ and $s_2$ throughout the detection chain}
}

\newglossaryentry{delta_Phi_0}{
	name=\ensuremath{\Delta\Phi_0},
	description={Part of \ensuremath{\Delta\Phi} that does not correspond to natural fringe frequency}
}

\newglossaryentry{delta_Phi_FS}{
	name=\ensuremath{\Delta\Phi_{\text{FS}}},
	description={Free-space phase difference}
}

\newglossaryentry{delta_Phi_LO}{
	name=\ensuremath{\Delta\Phi_{\text{LO}}},
	description={Phase difference between the Local Oscillators at the level of the infrared detectors}
}

\newglossaryentry{delta_Phi_RF}{
	name=\ensuremath{\Delta\Phi_{\text{RF}}},
	description={Radio-frequency phase difference}
}

\newglossaryentry{delta_Phi_corr}{
	name=\ensuremath{\Delta\Phi_{\text{corr}}},
	description={Correlator Phase difference}
}

\newglossaryentry{delta_tau}{
	name=\ensuremath{\Delta\tau},
	description={Total interferometric delay through the system}
}

\newglossaryentry{delta_tau_corr}{
	name=\ensuremath{\Delta\tau_{corr}},
	description={Delay introduced by the photonic correlator}
}

\newglossaryentry{delta_tau_FS}{
	name=\ensuremath{\Delta\tau_{\text{FS}}},
	description={Mid-infrared free-space delay}
}

\newglossaryentry{delta_tau_RF}{
	name=\ensuremath{\Delta\tau_{\text{RF}}},
	description={Delay introduced by difference of RF cable length}
}

\newglossaryentry{epsilon_k}{
	name=\ensuremath{\varepsilon_k},
	description={Equivalent RF constant input associated with the imbalance of intensity modulators \ensuremath{k}}
}

\newglossaryentry{E_LO}{
	name=\ensuremath{E_{\mathrm{LO}}},
	description={Electrical field of the local oscillator}
}

\newglossaryentry{E_AOFS_k}{
	name=\ensuremath{E_{\text{AOFS},k}},
	description={Electrical field at the output Acousto-Optic Frequency Shifter \ensuremath{k}}
}

\newglossaryentry{E_kZM_k}{
	name=\ensuremath{E_{\text{MZM},k}},
	description={Electrical field at the output of the Mach-Zender Modulator \ensuremath{k}}
}

\newglossaryentry{E_s}{
	name=\ensuremath{E_{s}},
	description={Electrical field of the science signal}
}

\newglossaryentry{E_pm}{
	name=\ensuremath{E_{\pm}},
	description={Electrical field at input \ensuremath{\pm} of the Balanced Photo-Diode}
}

\newglossaryentry{eta}{
	name=\ensuremath{\eta},
	description={Quantum efficiency}
}

\newglossaryentry{f_pm}{
	name=\ensuremath{f_{\pm}},
	description={Frequencies where the correlation signal is extracted at the output of the correlator}
}

\newglossaryentry{f12}{
	name=\ensuremath{f_{12}},
	description={Natural fringe frequency}
}

\newglossaryentry{f_corr}{
	name=\ensuremath{f_{\text{corr}}},
	description={Correlator response function to correlated inputs}
}

\newglossaryentry{f_noise}{
	name=\ensuremath{f_{\text{noise}}},
	description={Correlator response function to uncorrelated inputs}
}

\newglossaryentry{F_c}{
	name=\ensuremath{F_c},
	description={Coherent flux}
}

\newglossaryentry{F_c_nu}{
	name=\ensuremath{F_{c,\nu}},
	description={Spectral coherent flux for a single polarization}
}

\newglossaryentry{gamma_12}{
	name=\ensuremath{\gamma_{12}},
	description={Degree of coherence between \ensuremath{s_1} and \ensuremath{s_2}}
}

\newglossaryentry{G}{
	name=\ensuremath{G_{k}},
	description={Radio-Frequency gain on signal \ensuremath{s_k}}
}

\newglossaryentry{G_sys}{
	name=\ensuremath{G_{\text{sys}}},
	description={Total gain of the system}
}

\newglossaryentry{h_bar}{
	name=\ensuremath{\hbar},
	description={Reduced Planck constant}
}
\newglossaryentry{h_k}{
	name=\ensuremath{h_{k}},
	description={RF coherent signal at the detector output \ensuremath{k} or at the correlator input \ensuremath{k}}
}

\newglossaryentry{H_det}{
	name=\ensuremath{H_{\mathrm{det}}},
	description={Electrical transfer function of the detector}
}

\newglossaryentry{H}{
	name=\ensuremath{\mathcal{H}},
	description={Observable at the output of the photonic correlation}
}

\newglossaryentry{i}{
	name=\ensuremath{\mathbf{i}},
	description={Square root of $-1$}
}

\newglossaryentry{i_BPD}{
	name=\ensuremath{i_{\text{BPD}}},
	description={Signal at the output of the Balanced Photo-Diode}
}

\newglossaryentry{lambda}{
	name=\ensuremath{\lambda},
	description={Wavelength}
}

\newglossaryentry{n_F}{
	name=\ensuremath{n_F},
	description={Noise factor of the system relative to the quantum-noise}
}

\newglossaryentry{N}{
	name=\ensuremath{N},
	description={Number of measurements}
}

\newglossaryentry{N_lambda}{
	name=\ensuremath{N_\lambda},
	description={Number spectral channels}
}

\newglossaryentry{NEP_corr}{
	name=\ensuremath{\text{NEP}_{\text{corr}}},
	description={Noise Equivalent Power of the correlator}
}

\newglossaryentry{NEI_corr}{
	name=\ensuremath{\text{NEI}_{\text{corr}}},
	description={Noise Equivalent Input of the correlator}
}

\newglossaryentry{omega_LO}{
	name=\ensuremath{\omega_{\mathrm{LO},k}},
	description={Frequency of local oscillator \ensuremath{k}}
}

\newglossaryentry{omega_s}{
	name=\ensuremath{\omega_{s}},
	description={Frequency of the science signal}
}

\newglossaryentry{omega_0}{
	name=\ensuremath{\omega_{0}},
	description={Frequency of the fibered correlator laser}
}

\newglossaryentry{P_pm}{
	name=\ensuremath{\mathcal{P}_{\pm}},
	description={Correlation power at $f_{\pm}$}
}

\newglossaryentry{P_LO}{
	name=\ensuremath{P_{\text{LO}}},
	description={Optical power of the local oscillator}
}

\newglossaryentry{P_RF}{
	name=\ensuremath{P_{\text{RF},k}},
	description={Radio-frequency power of signal \ensuremath{s_k} after amplification}
}

\newglossaryentry{P_RF_coherent}{
	name=\ensuremath{P_{\text{RF,coherent}}},
	description={Radio-frequency equivalent of the coherent flux}
}

\newglossaryentry{P_RF_noise}{
	name=\ensuremath{P_{\text{RF,noise}}},
	description={Radio-frequency noise}
}

\newglossaryentry{P_s}{
	name=\ensuremath{P_s},
	description={Optical power of the science source}
}

\newglossaryentry{P_nu_s}{
	name=\ensuremath{P_{\nu,s}},
	description={Spectral optical power of the science source}
}

\newglossaryentry{phi_LO}{
	name=\ensuremath{\phi_{\mathrm{LO}}},
	description={Phase of the local oscillator}
}

\newglossaryentry{phi_s}{
	name=\ensuremath{\phi_{s}},
	description={Phase of the science signal}
}

\newglossaryentry{Phi}{
	name=\ensuremath{\Phi_{\text{astro}}},
	description={Astronomical phase}
}

\newglossaryentry{Phi_pm}{
	name=\ensuremath{\Phi_{\pm}},
	description={Phase at \ensuremath{\Delta\omega_{2}-\Delta\omega_{1}\pm2\pi f_{12}} frequencies from \ensuremath{\text{PSD}_{\text{corr}}}}
}

\newglossaryentry{PSD_background}{
	name=\ensuremath{\text{PSD}_{\text{background}}},
	description={Power spectrum Density of \ensuremath{i_{\text{BPD}}} at the output of the correlator in the absence of coherent signal}
}

\newglossaryentry{PSD_corr}{
	name=\ensuremath{\text{PSD}_{\text{corr}}},
	description={Power spectrum Density of \ensuremath{i_{\text{BPD}}} at the output of the correlator in the presence of the coherent signal}
}

\newglossaryentry{responsivity}{
	name=\ensuremath{\mathcal{R}_{k}},
	description={Detector \ensuremath{k} responsivity}
}

\newglossaryentry{s12}{
	name=\ensuremath{s_{k}},
	description={RF signal at the detector output \ensuremath{k} or at the correlator input \ensuremath{k}}
}

\newglossaryentry{sigma}{
	name=\ensuremath{\sigma},
	description={Standard deviation}
}

\newglossaryentry{time}{
	name=\ensuremath{t},
	description={Time}
}

\newglossaryentry{T_amp}{
	name=\ensuremath{T_{\text{amp}}},
	description={Noise temperature of the Radio-Frequency Amplifier}
}

\newglossaryentry{T_det}{
	name=\ensuremath{T_{\text{det}}},
	description={Noise temperature of the infrared detector}
}

\newglossaryentry{T_shot_noise}{
	name=\ensuremath{T_{\text{shot-noise}}},
	description={Noise equivalent temperature of the shot-noise}
}

\newglossaryentry{T_corr}{
	name=\ensuremath{T_{\text{corr}}},
	description={Noise equivalent temperature of the correlator noise floor}
}

\newglossaryentry{T_int}{
	name=\ensuremath{T_{\text{int}}},
	description={Integration time}
}

\newglossaryentry{V_pi}{
	name=\ensuremath{V_{\pi,k}},
	description={Half-wave voltage of the Mach-zender Modulator \ensuremath{k}}
}

\newglossaryentry{w_k}{
	name=\ensuremath{w_{k}},
	description={RF noise at the detector output \ensuremath{k} or at the correlator input \ensuremath{k}}
}

\newglossaryentry{Z}{
	name=\ensuremath{Z},
	description={Electrical impedance}
}

\newglossaryentry{Z_trans}{
	name=\ensuremath{Z_{\text{trans}}},
	description={Trans-impedance gain}
}

\newglossaryentry{ALMA}{
	name=ALMA,
	description={Atacama Large Millimeter/Submillimiter Array}
}

\newglossaryentry{AOFS}{
	name=AOFS,
	description={Acousto-Optic Frequency Shifter}
}

\newglossaryentry{ASE}{
	name=ASE,
	description={Amplified Spontaneous Emission}
}

\newglossaryentry{BPD}{
	name=BPD,
	description={Balanced Photo-Detector}
}

\newglossaryentry{CHARA}{
	name=CHARA,
	description={Center for High Angular Resolution Astronomy array}
}

\newglossaryentry{FWHM}{
	name=FWHM,
	description={Full Width Half Maximum}
}

\newglossaryentry{HIKE}{
	name=HIKE,
	description={Heterodyne Interferometric Kilometric Experiment}
}

\newglossaryentry{IR}{
	name=IR,
	description={Infrared}
}

\newglossaryentry{ISI}{
	name=ISI,
	description={Infrared Spatial Interferometer}
}

\newglossaryentry{LO}{
	name=LO,
	description={Local Oscillator}
}

\newglossaryentry{MZM}{
	name=MZM,
	description={Mach-zender Modulators}
}

\newglossaryentry{NEP}{
	name=NEP,
	description={Noise Equivalent Power}
}

\newglossaryentry{NEI}{
	name=NEI,
	description={Noise Equivalent Input}
}

\newglossaryentry{PFI}{
	name=PFI,
	description={Planet formation Imager}
}

\newglossaryentry{PLL}{
	name=PLL,
	description={Phase-Lock loop}
}

\newglossaryentry{QCD}{
	name=QCD,
	description={Quantum Cascade Detectors}
}

\newglossaryentry{QCL}{
	name=QCL,
	description={Quantum Cascade Laser}
}

\newglossaryentry{QWIP}{
	name=QWIP,
	description={Quantum Well Infrared Photodetectors}
}

\newglossaryentry{RF}{
	name=RF,
	description={Radio Frequency}
}

\newglossaryentry{SNR}{
	name=SNR,
	description={Signal to Noise Ratio}
}

\newglossaryentry{QE}{
	name=QE,
	description={Quantum efficiency}
}

\newglossaryentry{VLBI}{
	name=VLBI,
	description={Very Long Baseline Interferometry}
}

\newglossaryentry{VLTI}{
	name=VLTI,
	description={Very Large Telescope Interferometer}
}

\newcommand{\im}{\mathbf{i}}

\begin{document}
	\title{Proof of concept of heterodyne interferometry at $10.6~\mathrm{\mu m}$ using photonic correlation}
	
	\author{T. Allain$^{1,2,4}$, J.-P. Berger$^{1}$, G. Bourdarot$^{3}$, C. Sirtori$^{2}$ and H. Guillet de Chatellus$^{4}$
	}
	
	\date{\raggedright$^1$Univ. Grenoble Alpes, CNRS, IPAG, 38000 Grenoble, France\\
		$^2$Laboratoire de Physique de l’Ecole Normale Supérieure, ENS, Université PSL, CNRS, Sorbonne Université, Université de Paris, 24 rue Lhomond, 75005, Paris, France\\
		$^3$Max Planck Institute for extraterrestrial Physics, Giessenbach-straße 1, 85748 Garching, Germany\\
		$^4$Univ. Rennes, CNRS, Institut FOTON -- UMR 6082, 35000 Rennes, France\\
	}
	
	\maketitle
	
	\noindent\textbf{Keywords:} heterodyne, interferometry, mid-infrared, photonics, correlation
	
	\bigskip  
	
	\noindent Send correspondance to tituan.allain@univ-rennes.fr \& jean-philippe.berger@univ-grenoble-alpes.fr

   \maketitle

\abstract{
	Imaging complex dust environments such as the inner astronomical units of a planet forming disks requires dedicated mid-infrared interferometric facilities with kilometric baselines and a large number of telescopes. Extrapolating technologies from current facilities is not straightforward.
	We aim to demonstrate the feasibility of using mid-infrared heterodyne interferometry with a photonic correlation approach to recombine mid-infrared signals from distant telescopes. We want to determine what are the current technological limits of such system.
	We developed a laboratory demonstration bench that can correlate mid-infrared signals at $10~\mathrm{\mu m}$ with a photonic correlator. The photonic correlator uses commercially available telecom components at $1.5~\mathrm{\mu m}$ to transport and correlate heterodyne signals that could have up $10~\mathrm{GHz}$ bandwidth, directly extendable to $40~\mathrm{GHz}$. We used the demonstration bench to study the noise-levels and detection limits of a heterodyne interferometer with photonic correlation.
	We demonstrated the correlation of wideband mid-infrared signals with a photonic correlator. We characterized the performance of the system and analyzed the noise levels. We showed the photonic correlator is not limiting the detection and that it can be used to compensate free-space delay at $10~\mathrm{\mu m}$ with fiber delay at $1.5~\mathrm{\mu m}$. With our current sub-optimal commercial infrared detectors, we derived a detection limit of $130~\mathrm{Jy}$ coherent flux for $8~m$ class telescopes with $1~\mathrm{h}$ of incoherent integration. We discuss the possibility to lower the detection limit down to typical T-Tauri stars ($\sim\mathrm{Jy}$) using new detectors and coherent integration based upon local oscillator synchronization with telecom fiber links.}

\section{Introduction}

Aperture synthesis with Very Long Baseline interferometry (VLBI) and optical
infrared interferometry (e.g. Very Large Telescope Interferometer, VLTI or
CHARA) are currently the two techniques that provide the highest angular
resolution in astronomy. Optical interferometry, still confined at the few hundred meter scale (200 m maximum baseline for VLTI, 330 m for CHARA),
could be envisioned as one of the most promising techniques to go even further.
Its extension to a large number of telescopes and kilometric baselines would
represent a major step for observational astronomy by allowing imaging of
complex astrophysical targets. Nevertheless, such an infrastructure, as
proposed for exampleby the Planet Formation Imager (PFI) initiative (\cite{monnier2018})
will also require challenging technological developments that cannot be simply
extrapolated from existing ones (\cite{ireland2014,Ireland2016}).

Heterodyne detection offers a
complementary path to address the problem of kilometric baseline and aperture synthesis with a large number of telescopes because it is
less demanding on the infrastructure. In the past, through the pioneering
work of maser inventor and Nobel Prize  recipient C.H.  Townes and his team, heterodyne
detection was the first technique able to combine 2 and 3 telescopes in the
mid-infrared on the Infrared Spatial Interferometer (ISI) in UC Berkeley
(\cite{Hale2000}). ISI provided valuable scientific results well ahead of his
time, anticipating the following generation of direct mid-infrared
interferometric instruments such as MIDI and MATISSE. While this technique is
widely used in the radio regime its use in the optical has been limited to the
ISI experiment because of a fundamental quantum noise limitation.  Yet, the
mid-infrared regime ($\geq 10 ~\mu\rm{m}$) remains a domain where heterodyne can
be competitive with direct interferometry in particular when one envisions a
phased array of several tens of telescopes and narrow-band observations such as
PFI (\cite{Johnson2000}).

One of the critical aspects for improving the
sensitivity of heterodyne interferometry in the infrared is the ability to
detect signal of several THz. This requires very high bandwidth detectors
(several tens of GHz) and a multiplexing capability of several tens, if not
hundreds of spectral channels (\cite{ireland2014}). While such technologies are
still not operational, research in the field of sensing, telecommunication and
quantum technologies are pushing the boundaries in the right directions. In
particular, to overcome the limited speed ($<$3 GHz) of the HgCdTe detector, the
most advanced alternative for the 10-20 $\mu \rm{m}$ spectral range is provided
by photonics-enhanced Quantum Well Infrared Photodetectors (QWIP), and Quantum Cascade
Detectors (QCD) (\cite{Palaferri2018,Bigioli2020,Hakl2021,quinchard_high_2022,Lin2023})
where flat responses up to more than 70 GHz have been reported
(\cite{Hakl2021}). Other fast technologies have been proposed, namely using
graphene (\cite{Cakmakyapan2018}).

Such bandwidth requirements raise the question of the correlation of
the signals. The ALMA digital correlator is currently able to
correlate $8~\text{GHz}$ of bandwidth signals for 2016 telescope
pairs (\cite{escoffier2007}). Extrapolations of such a digital approach to the
mid-infrared would require both digitizing technologies and extremely
computing powers that are beyond current numerical capabilities (\cite{ireland2014}).

In our recent studies, we revived the idea of analog correlation
of heterodyne signals using photonics technologies.  Our approach was
based on the consideration that telecommunication industry provides,
off-the-shelf, all the building blocks requested for a high-bandwidth,
massively multiplexed correlation analog device. In previous
studies we focused our
efforts on the exploration of possible photonic architectures
(\cite{Bourdarot2020, Bourdarot2021, Bourdarot2022}). In this article
we demonstrate, in the laboratory, a complete proof-of-concept scheme
that handles the correlation of a two telescope mid-infrared
heterodyne interferometer using a near-infrared all-fibered
instrumental chain. In section \ref{sec:principle} we recall the
principle of photonic correlation and the underlying equations. In
section \ref{sec:setup} we describe our laboratory set-up and devote
section \ref{section:results} to the presentation of the main
measurements made with the bench. Section \ref{section:discussion} allows us
to discuss the performance and limitations of such an approach to
interferometric correlation.

%__________________________________________________________________

\section{Principles of photonic correlation}
\label{sec:principle}

\begin{figure}[t]
    \includegraphics[width=0.7\hsize]{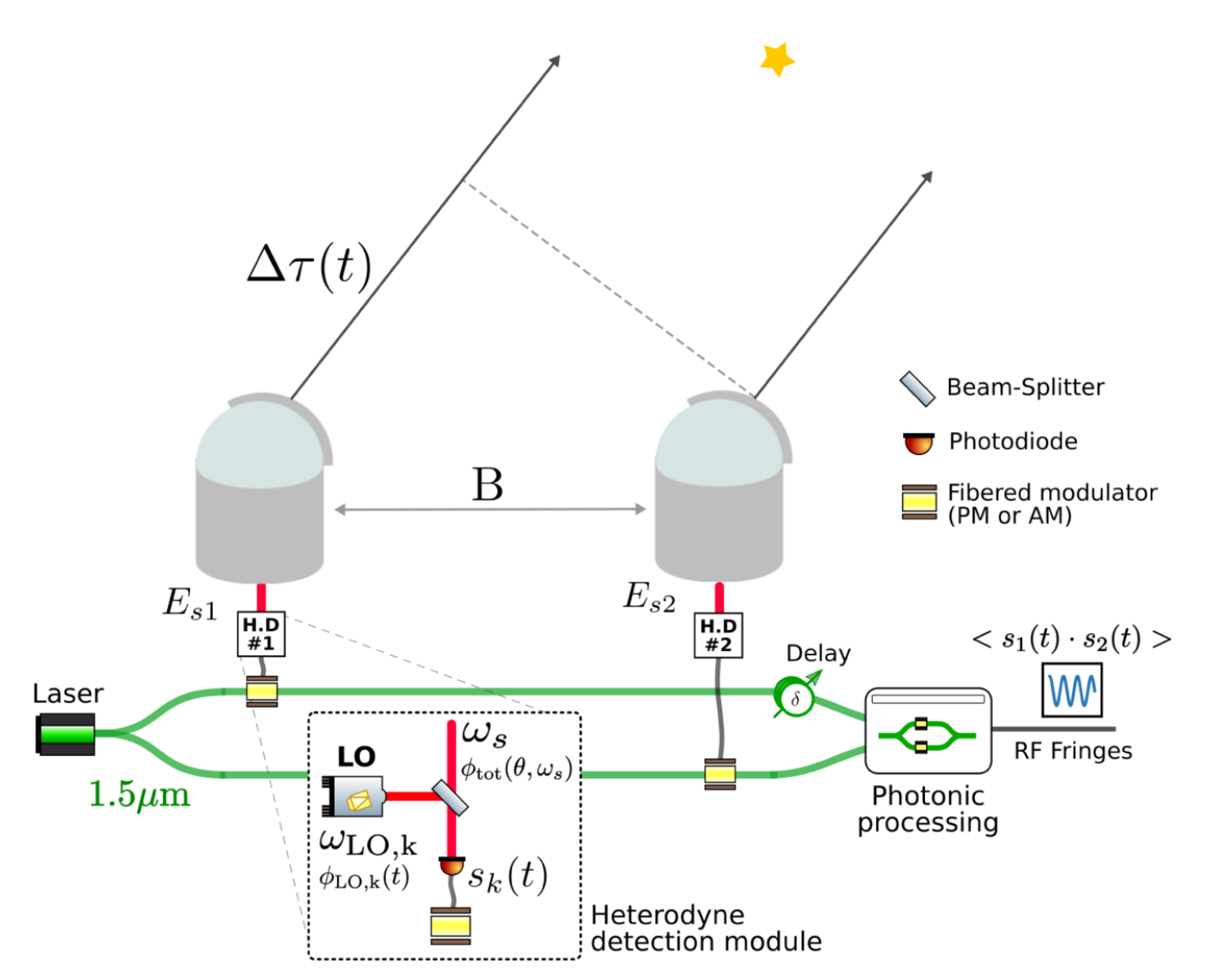}
    \caption{General scheme of a two telescope heterodyne interferometer, taken from \cite{Bourdarot2021}.}
    \label{fig:heterodyne_interferometry}
\end{figure}

\subsection{Heterodyne detection}

Heterodyne interferometry relies on the measurement of the electrical field at each telescope to retrieve the coherent flux $F_c$ of the astronomical signal and its astronomical phase $\Phi_{\text{astro}}$ at a given spatial frequency.

The astronomical optical field $E_s$ is collected by the telescopes and superimposed with local oscillators (LO) electrical field $E_{\mathrm{LO}}$, as illustrated on Fig. \ref{fig:heterodyne_interferometry}. In case of a monochromatic astronomical signal, $E_{\mathrm{LO}}$ and $E_s$ are given by equation (\ref{eq:electrical_fields}).

\begin{equation}
    \begin{aligned}
        &E_{\mathrm{LO}}=|E_{\mathrm{LO}}|e^{-\im\omega_{\mathrm{LO}}t-\im\phi_{\mathrm{LO}}}
        \\
        &E_{s}=|E_{s}|e^{-\im\omega_{s}t-\im\phi_{s}}
    \end{aligned}
	\label{eq:electrical_fields}
\end{equation}

A fast infrared detector is used to measure the beating (i.e. rapid fluctuations of the optical intensity) between the two fields are at an intermediate frequency (IF) $\omega_{if}=\omega_{\mathrm{LO}}-\omega_s$. The expression of the beating is given by  equation (\ref{eq:heterodyne_beating_monochromatic}). The expression of the electrical signal $h(t)$ that results from the detection of the intermediate frequency is given by equation (\ref{eq:heterodyne_detection}).

\begin{equation}
        |E_{\mathrm{LO}}+E_s|^2=|E_{\mathrm{LO}}|^2+|E_{s}|^2+2|E_{\mathrm{LO}}||E_{s}|\cos((\omega_s-\omega_{\mathrm{LO}})t+\phi_s-\phi_{\mathrm{LO}})
	\label{eq:heterodyne_beating_monochromatic}
\end{equation}

\begin{equation}
        \frac{h(t)}{Z\mathcal{R}}=P_{\text{LO}}+P_{s}+2\sqrt{P_{\text{LO}}P_{s}}H_{\mathrm{det}}(\omega_s-\omega_{\mathrm{LO}})\cos((\omega_s-\omega_{\mathrm{LO}})t+\phi_s-\phi_{\mathrm{LO}})
	\label{eq:heterodyne_detection}
\end{equation}

\noindent $P_{\text{LO}}$ is the optical power of the local oscillator of angular frequency $\omega_{\mathrm{LO}}$ and phase $\phi_{\mathrm{LO}}$; $P_{s}$ is the optical power of the  the astronomical signal of angular frequency $\omega_s$ and phase $\phi_s$; $\mathcal{R}$ is the responsivity of the detector, usually expressed in $[A/W]$; $H_{\mathrm{det}}$ is the normalised electrical transfer function of the detector; $Z$ is the impedance of the Radio-Frequency (RF) line ($50~\Omega$). Knowing the values $P_{\text{LO}}$, $\phi_{\mathrm{LO}}$, $\mathcal{R}$ and $H_{\mathrm{det}}(\omega_s-\omega_{\mathrm{LO}})$, we can retrieve the measurement of the astronomical signal electrical field $E_s\propto\sqrt{F_c}\cos(\omega_st+\phi_s)$ at the intermediate frequency.

For a wideband astronomical source, the electrical signal consists in the sum of the intermediate frequencies from the beating of the local oscillator with the spectral components of the astronomical signal. The larger the electrical bandwidth of the detector, the broader the intermediate frequencies span is.

\subsection{Signal correlation}

Once the electric field of the astronomical signal is measured using heterodyne detection at each telescope, resulting in two signals $h_1$ and $h_2$, we perform the correlation of the signal which consists in measuring the average of the product of the signals $\langle h_1(t)h_2(t-\Delta\tau)\rangle_{T_{\text{int}}}$ over integration time $T_{\text{int}}$. We introduced a delay $\Delta\tau$ using equation (\ref{eq:tau}) as the sum of all the delays that are introduced between the two arms of the interferometer throughout the detection chain: free-space delay $\Delta\tau_{\text{FS}}$, RF delay $\Delta\tau_{\text{RF}}$ and correlator delay $\Delta\tau_{corr}$.

\begin{equation}
	\Delta\tau=\Delta\tau_{\text{FS}}+\Delta\tau_{\text{RF}}+\Delta\tau_{corr}
	\label{eq:tau}
\end{equation}

The correlation product $\langle h_1(t)h_2(t-\Delta\tau)\rangle_{T_{\text{int}}}$ is proportional to $F_c e^{\mathbf{i}\Phi_{\text{astro}}}$ where $F_{c}$ is the coherent flux of the astronomical source. Therefore, it carries the interferometric observables of the astronomical object (\cite{Bourdarot2021}). 

The shot-noise limited Signal-to-Noise Ratio (SNR) for the correlation of heterodyne signals is given by
equation (\ref{eq:SNR_correlation_theoretical}). It scales as the square
root of the detector bandwidth $\Delta\nu_{\text{det}}$ and integration time
$T_{\text{int}}$. It should be noted that the SNR does not depend on the LO
power $P_{\text{LO}}$. However, increasing the LO power will enable reaching
that shot-noise limited regime since $P_{\text{LO}}\gg2\sqrt{P_{\text{LO}}P_s}\gg P_s$.

\begin{equation}
	\text{SNR}=\frac{F_{c,\nu}}{\hbar\omega_{\mathrm{LO}}}\eta\sqrt{\Delta\nu_{\mathrm{det}}2T_{\text{int}}}
	\label{eq:SNR_correlation_theoretical}
\end{equation}

\noindent Where $F_{c,\nu}$ is the spectral coherent flux of the astronomical source for a single polarization (in $[W/Hz]$); $\eta$ is the quantum efficiency of the infrared detectors; $\Delta\nu_{\mathrm{det}}$ is the equivalent bandwidth of the detectors; $T_{\text{int}}$ is the integration time; $e$ is the elementary charge; $\hbar$ is the reduced Planck constant.

Similarly to direct interferometry which uses optical delay lines to
co-phase the telescopes, heterodyne interferometry requires delay
compensation between $h_1$ and $h_2$ to retrieve the astronomical
observables from the correlation product
$\langle h_1(t)h_2(t)\rangle_{T_{\text{int}}}$. The delay compensation
can be performed with hardware to ensure stable zero delay
$\Delta\tau=0$ between the different arms of the interferometer when
performing the correlation. Interferometric delay compensation in a
heterodyne interferometer with a photonic correlator can be done with
$1.5~\mathrm{\mu m}$ off-the-shelf fibered telecom delay-lines which
would be more handy than bulky $10~\mathrm{\mu m}$ free-space delay
lines. In digital correlators, the delay can be \textit{a posteriori} compensated.

Additionally, since the local oscillator phase variations are indistinguishable from astronomical field phase variations, heterodyne interferometry requires a stable phase-locking of the local oscillators between each telescope. 

\subsection{Photonic correlation}

\label{p:photonic_correlation}

\begin{figure}[t]
    \includegraphics[width=0.7\hsize]{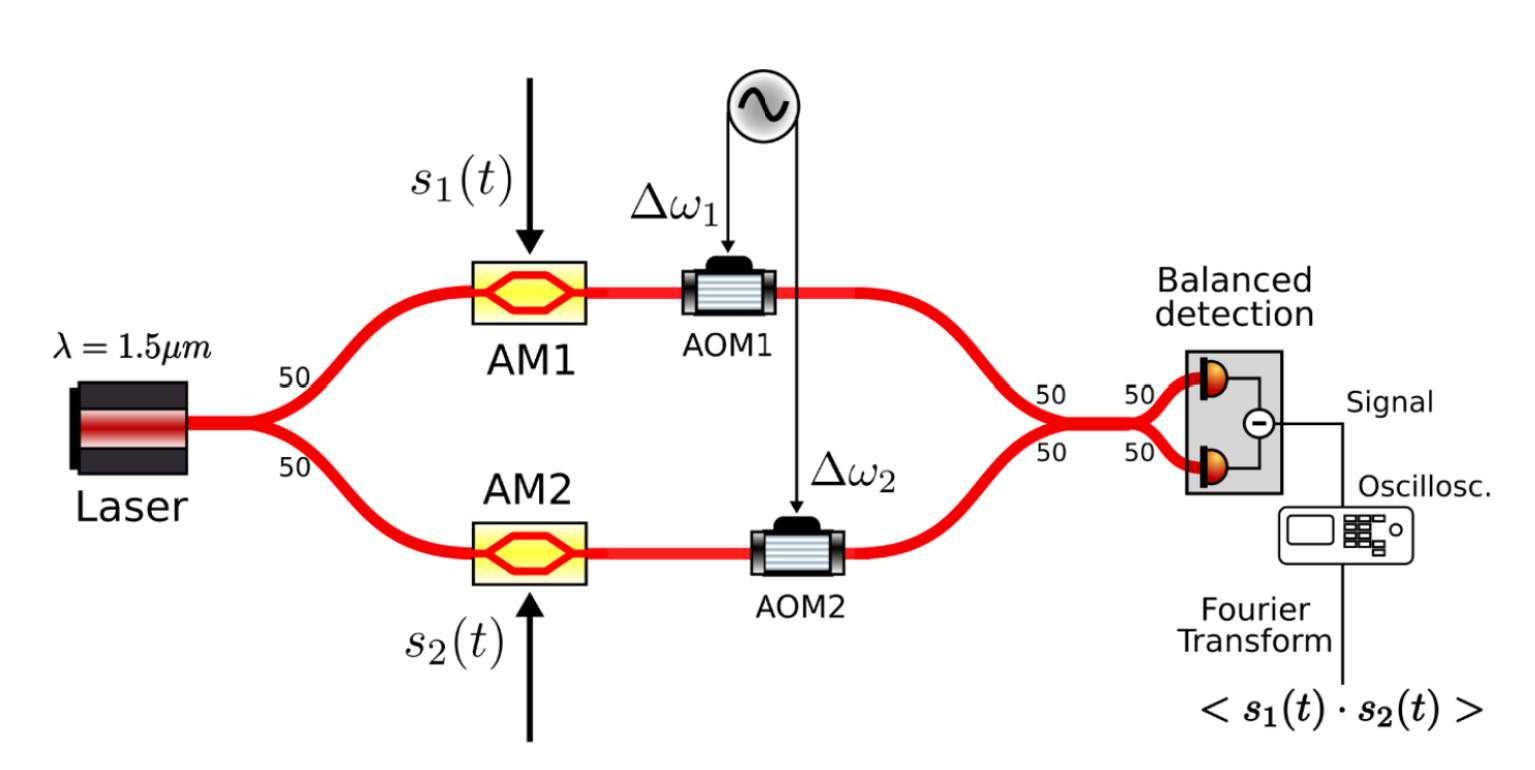}
    \caption{Photonic correlation scheme, taken from \cite{Bourdarot2021}. The Amplitude Modulators (AM) are Mach-Zehnder intensity Modulators. The frequency shift is performed with Acousto-Optic Modulators (AOM, also called AOFS).}
    \label{fig:photonic_correlator}
\end{figure}

The principle of amplitude photonic correlation was previously
described by \cite{Bourdarot2021} and is illustrated on
Fig. \ref{fig:photonic_correlator}. We remind here the main equations describing how it operates. A more complex analytical model that includes the noise terms added by the components in the correlator (notably the intensity modulators) can be found in \cite{allain2024}. A continuous fibered telecom laser
at $\omega_0$ is split in two channels\footnote{It is not necessary to
  have a single laser for both channels, two different lasers could be
  used with a phase-lock-loop to ensure stable phase difference and
  frequency difference. In that case, Acousto-Optic Frequency Shifters
  would not be required.}. The $h_1$ and $h_2$ heterodyne signals are
encoded in their respective channels using Amplitude Modulators (AM). This is done with fibered Mach-Zehnder intensity modulators (MZM), biased at their minima of transmission. The electrical fields $E_{\text{MZM},k}$ ($k$
is used to designate either channel $1$ or channel $2$) at the output
of the amplitude modulators are given by equation (\ref{eq:MZM}) at
first order in $\beta_kh_k(t)$.

\begin{equation}
	E_{\text{MZM},k}(t)=e^{-\im\omega_{0}t}\left(1+e^{\im\pi+\im\beta_k \sqrt{G_k}h_k(t)}\right)
	\simeq-\im e^{-\im\omega_{0}t}\beta_k \sqrt{G_k}h_k(t)
	\label{eq:MZM}
\end{equation}

\noindent $G_k$ is the RF power gain on signal $h_k$. The $\beta_k$ constants are equal to $\pi/V_{\pi,k}$ where  $V_{\pi,k}$ represents the signal that should be applied on the modulator to obtain a phase delay of $\pi$ between the two arms of the Mach-Zehnder.

The laser frequencies on each arm are shifted by Acousto-Optic Frequency Shifters (AOFS) by $\Delta\omega_1$ and $\Delta\omega_2$, resulting in a frequency difference $\Delta\omega_2-\Delta\omega_1$ between the two arms. The electrical fields $E_{\text{AOFS},k}$ at the output of the AOFSs are given by equation (\ref{eq:AOFS}). 
 
\begin{equation}
    E_{\text{AOFS},k}(t)=E_{\text{MZM},k}(t)e^{-\im\Delta\omega_kt}=e^{-\im(\omega_{0}+\Delta\omega_k)t}\beta_k \sqrt{G_k} h_k(t) 
	\label{eq:AOFS}
\end{equation}
 
The $E_{\text{AOFS},k}$ fields are recombined using a $50/50$ fiber coupler, resulting of two fields $E_+$ and $E_-$. For the sake of simplicity, we assume that the transmitted optical power is equally distributed between the two channels. The $E_{\pm}$ fields are detected by a balanced photodetector (BPD), leading to the measurement of $|E_{\pm}|^2$ given by equation (\ref{eq:I_+-}) where $\Delta\phi_{\text{corr}}$ is the phase difference of the fibered laser between the two arms of the correlator. 

\begin{equation}
	\begin{aligned}
		|E_{\pm}(t)|^2=&|E_{\text{AOFS},1}(t) \pm E_{\text{AOFS},2}(t-\Delta\tau)|^2
		\\
		=&\left|e^{-\im(\omega_{0}+\Delta\omega_1)t}\beta_1\sqrt{G_1}h_1(t)\pm e^{-\im(\omega_{0}+\Delta\omega_2)t-\im\Delta\phi_{\text{corr}}}\beta_2 \sqrt{G_2}h_2(t-\Delta\tau)\right|^2
		\\
		=&\beta_1^2G_1h_1^2(t)+\beta_2^2G_2h_2^2(t-\Delta\tau)pm2\beta_1\beta_2\cos((\Delta\omega_{2}-\Delta\omega_{1})t+\Delta\phi_{\text{corr}})\sqrt{G_1G_2}h_1(t)h_2(t-\Delta\tau)
	\end{aligned}
	\label{eq:I_+-}
\end{equation}

The balanced photodetector output $i_{\text{BPD}}(t)$ reads the difference between $|E_{+}|^2$ and $|E_{-}|^2$ which is given by equation (\ref{eq:BPD}) \footnote{The averaging of the $h_1(t)h_2(t-\Delta\tau)$ product originates from the response time of the balanced photodiode}.

\begin{equation}
	\begin{aligned}
		i_{\text{BPD}}(t)=&|E_{+}(t)|^2-|E_{-}(t)|^2
		\\
		=&4\beta_1\beta_2\cos((\Delta\omega_{2}-\Delta\omega_{1})t+\Delta\phi(t)+\Phi_{\text{astro}})\sqrt{G_1G_2} |\langle h_1(t)h_2(t-\Delta\tau)\rangle|
	\end{aligned}
	\label{eq:BPD}
\end{equation}

\noindent Where $\Delta\Phi(t)$ is the total accumulated phase difference between $h_1$ and $h_2$ throughout the detection chain, which is given by equation (\ref{eq:delta_phi}): $\Delta\Phi_{\mathrm{LO}}(t)$ is the phase difference between the local oscillators at each infrared detector, $\Delta\Phi_{\mathrm{FS}}(t)$ is the phase difference from the free space propagation, $\Delta\Phi_{\mathrm{RF}}(t)$ is the phase difference in the RF link and $\Delta\Phi_{\mathrm{corr}}(t)$ is the phase difference in the photonic correlator.

\begin{equation}
    \Delta\phi(t)=\Delta\phi_{\text{FS}}(t)+\Delta\phi_{\mathrm{LO}}(t)+\phi_{\text{RF}}(t)+\Delta\phi_{\text{corr}}(t)
    \label{eq:delta_phi}
\end{equation}

From $i_{\text{BPD}}(t)$, we can extract the correlation product $\langle h_1(t)h_2(t-\Delta\tau)\rangle$ carrying $F_c$ and $\Phi_{\text{astro}}$. We can re-write $i_{\text{BPD}}$ into equation (\ref{eq:s1_s2}).

\begin{equation}
    \begin{aligned}
        i_{\text{BPD}}(t)=&F_cG_{\text{sys}}\cos((\Delta\omega_{2}-\Delta\omega_{1})t+\Delta\phi(t)+\Phi_{\text{astro}})
    \end{aligned}
    \label{eq:s1_s2}
\end{equation}

\noindent Knowing the complex gain of system $G_{\text{sys}}e^{\im\Delta\phi(t)}$, the coherent flux $F_c$ and the astronomical phase $\Phi_{\text{astro}}$ can be retrieved from the $\Delta\omega_{2}-\Delta\omega_{1}$ frequency component of $i_{\text{BPD}}(t)$.

\subsection{Natural fringe frequency}

For an on-sky heterodyne interferometer, the free-space delay between the telescopes $\Delta\tau_{\text{FS}}$ changes with time due to the Earth rotation, which changes the astronomical object position in the sky. The total phase term $\Delta\Phi$ can be written in the form of equation (\ref{eq:delta_phi_natural_freq}).  

\begin{equation}
    \Delta\phi(t)=\Delta\phi_{0}(t)+\Delta\phi_{\text{corr}}(t)+2\pi f_{12}t
    \label{eq:delta_phi_natural_freq}
\end{equation}

\noindent $f_{12}$ is referred to as the fringe frequency. It is predictable and results from two contributions: the rotation of the Earth and the frequency difference between the local oscillators. For a given baseline $\vec{B}$, the rotation of the Earth impinges a maximum fringe frequency $f_{12}= \lVert \vec{B} \rVert/\lambda\times (7.27\times 10^{-5}~\text{rad/s})$ (\cite{Thompson2017}) ($\sim 1~\text{kHz}$ for a $1~\text{km}$ baseline). The frequency difference between the local oscillators can be tuned to adjust $f_{12}$ to an arbitrary value.

In the presence of the $f_{12}$ natural frequency term, the $i_{\text{BPD}}$ signal from equation (\ref{eq:BPD}) becomes equation (\ref{eq:BPD_natural_frequency}).

\begin{equation}
	\begin{aligned}
		i_{\text{BPD}}(t)\propto\frac{1}{2}F_cG_{\text{sys}}\Big(
		&\cos(2\pi f_-t+\Phi_{-})+\cos(2\pi f_+t+\Phi_{+})\Big)
	\end{aligned}
	\label{eq:BPD_natural_frequency}
\end{equation}

\noindent Where:
\begin{equation}
	f_\pm=\Delta\omega_{2}/2\pi-\Delta\omega_{1}/2\pi\pm f_{12}
	\label{eq:f_pm}
\end{equation}
\begin{equation}
	\Phi_{\pm}(t)=\Delta\phi_{\text{corr}}(t)\pm\Delta\Phi_0(t)\pm\Phi_{\text{astro}}
	\label{eq:phi_pm}
\end{equation}

The coherent flux is now encoded at both the $f_-$ and $f_+$ frequencies at the output of the balanced photodiode. The coherent flux can therefore be extracted by measuring the power of the $f_-$ and $f_+$ frequency components of $i_{\text{BPD}}(t)$.

To extract the astronomical phase $\Phi_{\text{astro}}$, the instrumental phase terms must be stabilized during the integration time. The $\Delta\Phi_0(t)+\Phi_{\text{astro}}$ sum can be extracted by subtracting  $\Phi_+$ and $\Phi_-$ which cancels the phase term from the correlator ($\Delta\phi_{\text{corr}}$). Knowledge of the $\Delta\Phi_0$ value is necessary to retrieve the absolute value of the astronomical phase. If $\Delta\Phi_0$ is unknown, only the relative value of $\Phi_{\text{astro}}$ is accessible which can still be used for astronomical measurements with closure-phase (\cite{hale2003}).

\subsection{Correlation in the presence of noise}

The correlator performs the multiplication of the RF input signals. These input signals contain noise terms that we write $w_1$ and $w_2$ such that the inputs are $s_1=h_1+w_1$ and $s_2=h_2+w_2$. The bandwidths of $h_1$, $h_2$, $s_1$ and $s_2$ are determined by the infrared detectors' ones. ($\Delta\nu_{\mathrm{det}}$).
 
An ideal correlator performs the multiplication of $s_1$ and $s_2$ (equation (\ref{eq:noise_product})) in which we recover the $h_1h_2$ product that encodes the astronomical coherent flux and additional noise terms ($h_1w_2$, $h_2w_1$ and $w_1w_2$). 

\begin{equation}
    s_1s_2=(h_1+w_1)(h_2+w_2)=h_1h_2+h_1w_2+h_2w_1+w_1w_2
    \label{eq:noise_product}
\end{equation}

Typical astronomical observations, with faint sources, correspond to a low-signal regime where the noise from so-called cross-terms $h_1w_2$ and $h_2w_1$, and from the correlation term $h_1h_2$ are negligible compared with the $w_1w_2$ term (\cite{anantharamaiah1989}). Therefore, we ignore the $h_1w_2$, $h_2w_1$ and $h_1h_2$ noise contributions in the following description\footnote{In addition to the correlation signal, $h_1h_2$ contains a wide-band contribution originating from the product of $h_1$ and $h_2$. This wide-band term contributes to the noise floor.}.

The $w_1w_2$ term is the product of two noise signals. The product of two white noise signals with bandwidth $\Delta\nu$ yields a noise signal whose spectrum decreases linearly to zero at $2\Delta\nu_{\text{det}}$ (\cite{Kish2012}). Thus, despite $w_1$ and $w_2$ not being correlated, a fraction of the $w_1w_2$ power leaks to the $f_-$ and $f_+$ frequencies and affects the measurement of the coherent flux.

\begin{figure}
    \includegraphics[width=0.7\hsize]{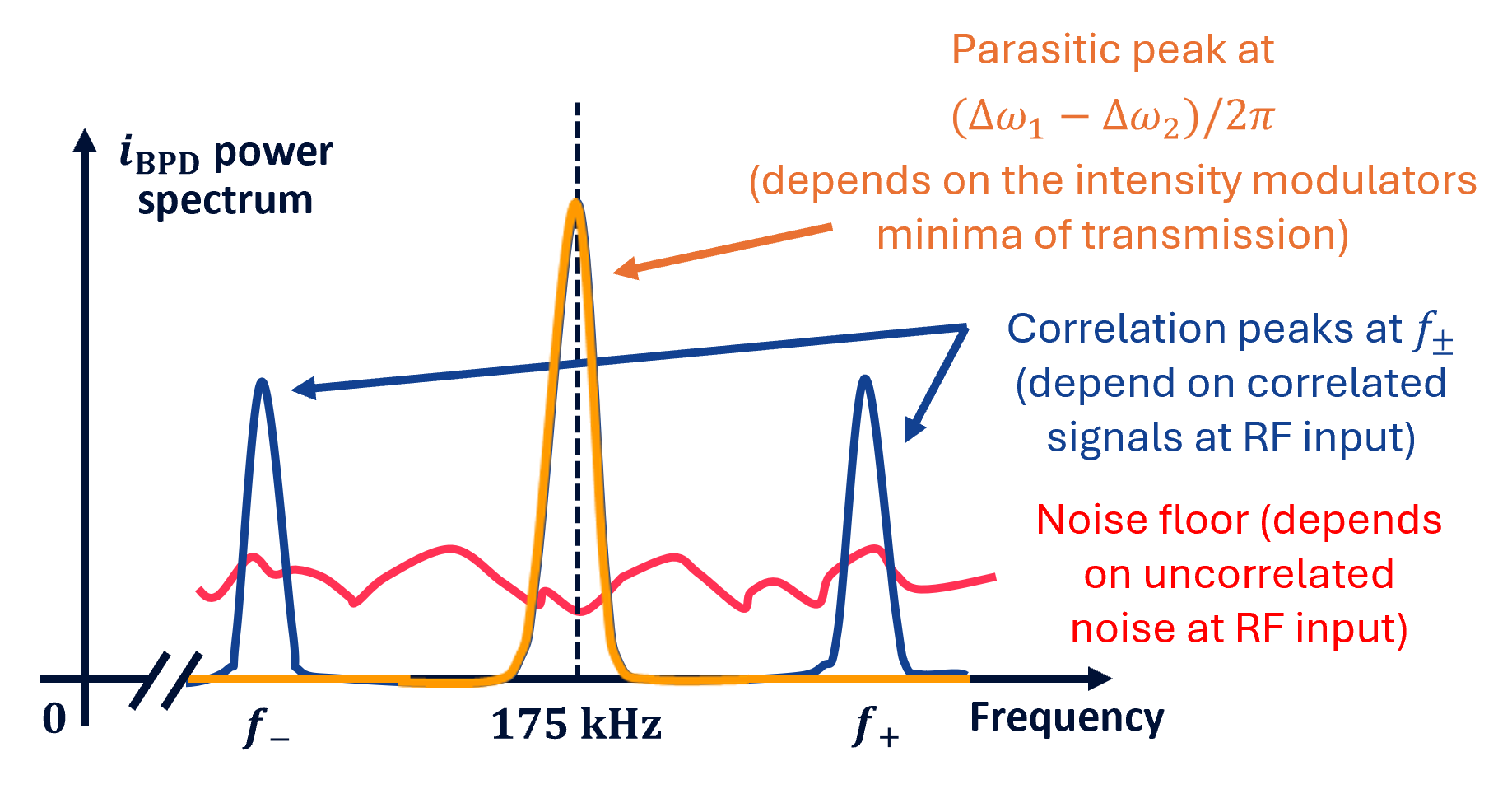}
    \caption{Typical shape of the $i_{\text{BPD}}$ power spectrum in the presence of with contributions and imperfections.}
    \label{fig:correlation_output}
\end{figure}

A schematic power spectrum of $i_\text{BPD}$ is shown on Fig. \ref{fig:correlation_output}. Coherent signals $h_1$ and $h_2$ contribute to the $f_-$ and $f_+$ peaks. The width of the peaks is set by the stability of $f_{12}$ and $\Delta\omega_{2}-\Delta\omega_{1}$. If they are stable enough, then the peaks are Fourier transform limited: the larger the integration time, the narrower the peaks. Despite carrying no coherence, uncorrelated noise terms $w_1$ and $w_2$ contribute to increasing the noise floor of the correlator. Additionally, some imperfection in the correlation, mainly the non-infinite extinction ratio of the Mach-Zehnder modulators lead to the presence of a peak at $\Delta\omega_{2}-\Delta\omega_{1}$. 

\subsection{Signal-to-noise ratio}

We define our main observable $\mathcal{H}$ as the sum of the correlation powers $\mathcal{P}_{-}$ and $\mathcal{P}_{+}$ that can be extracted at $f_-$ and $f_+$ from the power spectrum of $i_{\text{BPD}}$ (equation (\ref{eq:P_corr})). Careful calibration of the overall set-up will enable us to retrieve the coherent flux $F_c$ of the science source from the value of $\mathcal{H}$.

\begin{equation}
    \mathcal{H}=\mathcal{P}_{-}+\mathcal{P}_{+}
    \label{eq:P_corr}
\end{equation}

Rigorously, we can write $\mathcal{P}_{-}$ and $\mathcal{P}_{+}$ as the sum of two functions $f_{\text{corr}}$ and $f_{\text{noise}}$ (equation (\ref{eq:correlator_calibration}))\footnote{Rigorously, we should differentiate the $f_{\text{corr}}$ and $f_{\text{noise}}$ functions for $\mathcal{P}_{-}$ and $\mathcal{P}_{+}$ but, in practice, the measured functions are virtually the same.}. The $f_{\text{corr}}$ function carries the signature of the coherent flux $F_c$. The $f_{\text{noise}}$ function carries all the noise sources from the system.

\begin{equation}
    \mathcal{P}_{\pm} =f_{\text{corr}}\left(F_c\right)+f_{\text{noise}}
	\label{eq:correlator_calibration}
\end{equation}

From equation (\ref{eq:BPD_natural_frequency}), we know that $f_{\text{corr}}$ can be written in the form of equation (\ref{eq:f_corr}).

\begin{equation}
    f_{\text{corr}} =\alpha_{\text{IR}}\alpha_{\text{corr}} F_c^2=\alpha_{\text{corr}}{P_{\text{RF,coherent}}}^2
	\label{eq:f_corr}
\end{equation}

\noindent Where $\alpha_{\text{IR}}$ is a constant that depends on the infrared detection stage (detector responsivities, local oscillator power, local oscillator and signal mode coupling, etc), and $\alpha_{\text{corr}}$ is a constant that depends on physical parameters of the correlator's components (fiber transmission coefficients, amplitude modulator responses, etc). $P_{\text{RF,coherent}}=\sqrt{\alpha_{\text{IR}}}F_c$ is a proxy for the coherent flux $F_c$, but in the RF domain, after the mid-infrared detection.

%The exact derivation of the $f_{\text{noise}}$ is out of the scope of this article.
It can be shown that, if we place ourselves in the weak signal regime (the amplitudes of $h_1$ and $h_2$ are much lower than the amplitudes of $w_1$ and $w_2$), and if we suppose the RF noises $w_1$ and $w_2$ have similar amplitudes, then $f_{\text{noise}}$ can be written in the form of a a second order polynomial (equation (\ref{eq:f_noise})) \cite{allain2024}.

\begin{equation}
    f_{\text{noise}} \simeq \alpha_{\text{IR}}\left(A+B\times P_{\text{RF,noise}}+\frac{C}{\Delta\nu_{\text{det}}}\times{P_{\text{RF,noise}}}^2\right)
	\label{eq:f_noise}
\end{equation}

\noindent Where $P_{\text{RF,noise}}$ is the RF power of $w_1$ and $w_2$; $A$, $B$ and $C$ are constants that depend on the physical parameters of the correlator. In an ideal correlator, we should have $A=0$ and $B=0$, only keeping the quadratic noise terms ($C$). $A$ represents the correlator noise floor which depends on the noise sources inside the correlator (bias control of the Mach-Zehnder modulators, RF driving of the AOFS, detector noise, ...). The $B$ term originates from the correlation between noise terms and a constant signal: in a real system, the presence of even residual imbalance between the arms of the Mach-Zehnder modulators is equivalent to adding constants terms $\varepsilon_k$ to the RF input signals ($s_k=\varepsilon_k+h_k+w_k$) and thus leads to $B\neq0$. Last, the $C$ term accounts for the unavoidable $w_1w_2$ noise product in equation (\ref{eq:noise_product}). The $\Delta\nu_{\text{det}}$ dependence comes from the previously mentioned spread of the $w_1w_2$ noise product from 0 to $2\Delta\nu_{\text{det}}$. The imperfections in the system, notably from the intensity modulators and the AOFSs also push $C$ beyond the theoretical limit, but their contributions are negligible ($<0.1\%$ increase). Higher order terms are negligible. 

Using similar formalism to the one in \cite{Thompson2017} (6.2), it can be shown that the Signal-to-Noise Ratio (SNR) on the measurement of $F_c$ from the measurement of $\mathcal{P}_{-}$ or $\mathcal{P}_{+}$ is $\sqrt{f_{\text{corr}}/f_{\text{noise}}}$. Therefore, assuming the noises from $\mathcal{P}_{-}$ and $\mathcal{P}_{+}$ are independent, we can deduce that the SNR on $F_c$ from the measurement of $\mathcal{H}=\mathcal{P}_{-}+\mathcal{P}_{+}$ is given by equation (\ref{eq:SNR_gamma}).

\begin{equation}
    \operatorname{SNR(F_{c})} =\sqrt{2\frac{f_{\text{corr}}(F_c)}{f_{\text{noise}}}}
	\label{eq:SNR_gamma}
\end{equation}

\section{Description of the set-up}
\label{sec:setup}

\begin{figure*}
    \includegraphics[width=1\hsize]{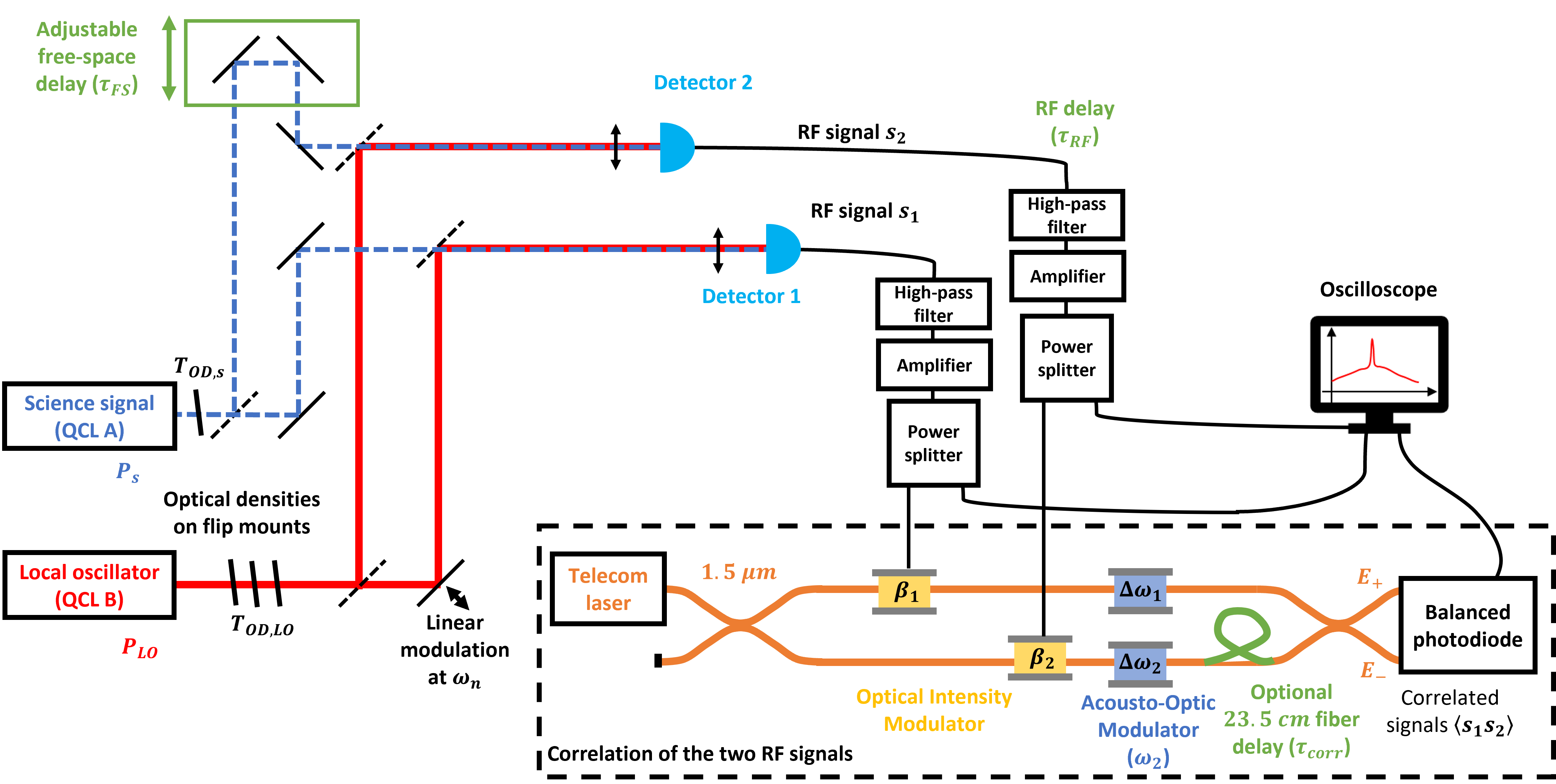}
    \caption{Representation of the current state of the demonstration bench.}
    \label{fig:bench}
\end{figure*}

\begin{figure*}
    \centering
    \includegraphics[width=1\hsize]{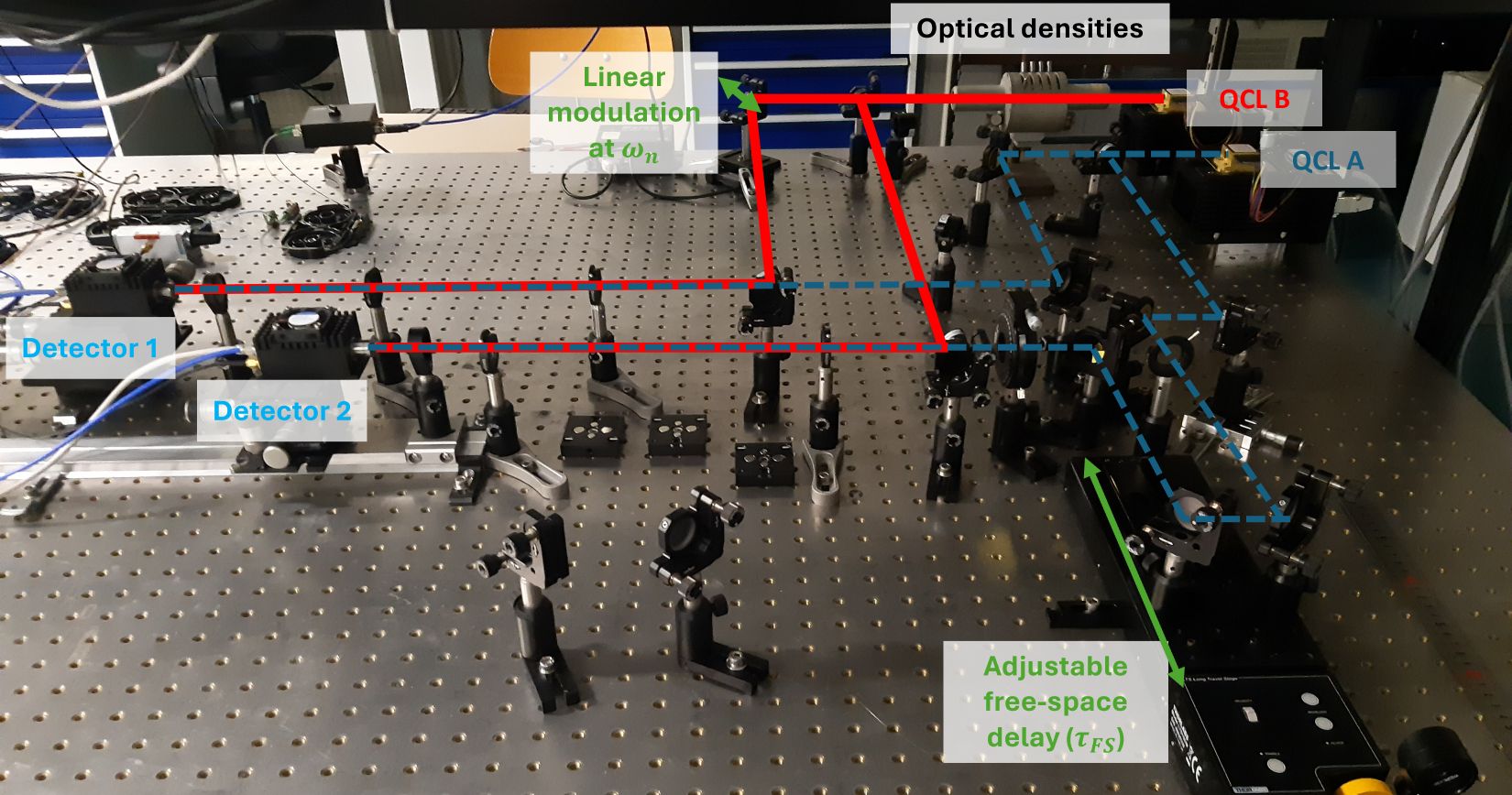}
    \caption{Photograph of the IR heterodyne detection stage of the bench with annotations highlighting the main elements.}
    \label{fig:bench_photo}
\end{figure*}

The demonstration set-up can be divided into two parts: the mid-infrared heterodyne detection part at $10.6~\mathrm{\mu m}$ and the photonic correlation part at $1.5~\mathrm{\mu m}$. Fig. \ref{fig:bench} illustrates the overall set-up. A picture of the optical table can be seen in Fig. \ref{fig:bench_photo}.

\subsection{Mid-infrared heterodyne detection stage}

The mid-infrared heterodyne detection is organised as follows. A Quantum Cascade Laser (QCL) is focused on two off-the-shelf \textit{VIGO} infrared detectors 1 and 2 with electronic bandwidths of roughly $900~\text{MHz}$. This QCL acts as a local oscillator. A modulation of the path between the local oscillator and detector 1 is applied by moving a mirror back and forth at $0.7~\text{mm/s}$ velocity. This results in a Doppler-shifted frequency $f_{12}=\pm 70~\text{Hz}$ between detector 1 and detector 2. 
Another QCL is used as a science source and is superimposed with the local oscillator onto the two detectors. The driving electrical currents of the QCLs can be adjusted to change their frequencies. We use optical densities to decrease the power of the local oscillator when needed. Stability on the order of $\simeq 10~\text{MHz}$ can be achieved on their relative frequency difference in open loop.

The science source QCL can be set below threshold in the Amplified Spontaneous Emission (ASE) regime in which it emits a weak ($\sim 1-10~\text{nW}$) signal with an optical bandwidth on the order of ($\sim 10-100~\text{MHz}$) (see Appendix~\ref{app:QCL}). The main advantage of using QCL in an ASE regime is the ability to align and calibrate the system in laser mode before switching to the larger and weaker ASE source. The larger the spectrum, the shorter the coherence length of the source. The typical shape of the emitted spectrum of the ASE can be seen in Appendix \ref{app:QCL} (Fig.~\ref{fig:QCL}).

A delay line allows us to set the optical path difference between detectors 1 and 2 up to $95~\text{cm}$ on the science signal.

\subsection{$1550~nm$ photonic correlator}

The photonic correlator follows the architecture presented in \ref{p:photonic_correlation}. The correlator performs the correlation of the wideband signals $s_1$ and $s_2$ from the mid-infrared detectors. $s_1$ and $s_2$ are filtered with $90~\text{MHz}$ high-pass filters and amplified with either one or two $+23~\text{dB}$ RF amplifiers. Filtering is used to improve the robustness of the correlator to low-frequency sources of noise in the system and to curb the linear noises terms of the correlator (the $B$ term in equation (\ref{eq:f_noise})). The choice of amplification level will be discussed in Section \ref{section:discussion}. 

Signals $s_1$ and $s_2$ are encoded on the arms of the correlator
using commercial Mach-Zehnder intensity modulator with $>10~\text{GHz}$
bandwidths\footnote{One of our Mach-Zehnder intensity modulator has a bandwidth limitation of $10~\text{GHz}$ while the other has a bandwidth of  $20~\text{GHz}$, similar $40~\text{GHz}$ modulators being available. }.  
A bias is applied in closed loop to the modulators to actively stabilize them at their minima of transmission. The two arms of the correlator are then shifted in frequency using acousto-optic frequency shifters that are respectively driven at $80.000$ and $80.175~\text{MHz}$. The two arms of the correlator are recombined using a $50/50$ fiber coupler whose outputs are sent to a commercial balanced photodetector with adjustable gain and electrical bandwidth.

\begin{figure*}
	\centering
	\begin{minipage}{0.49\hsize}
		\includegraphics[width=1\linewidth]{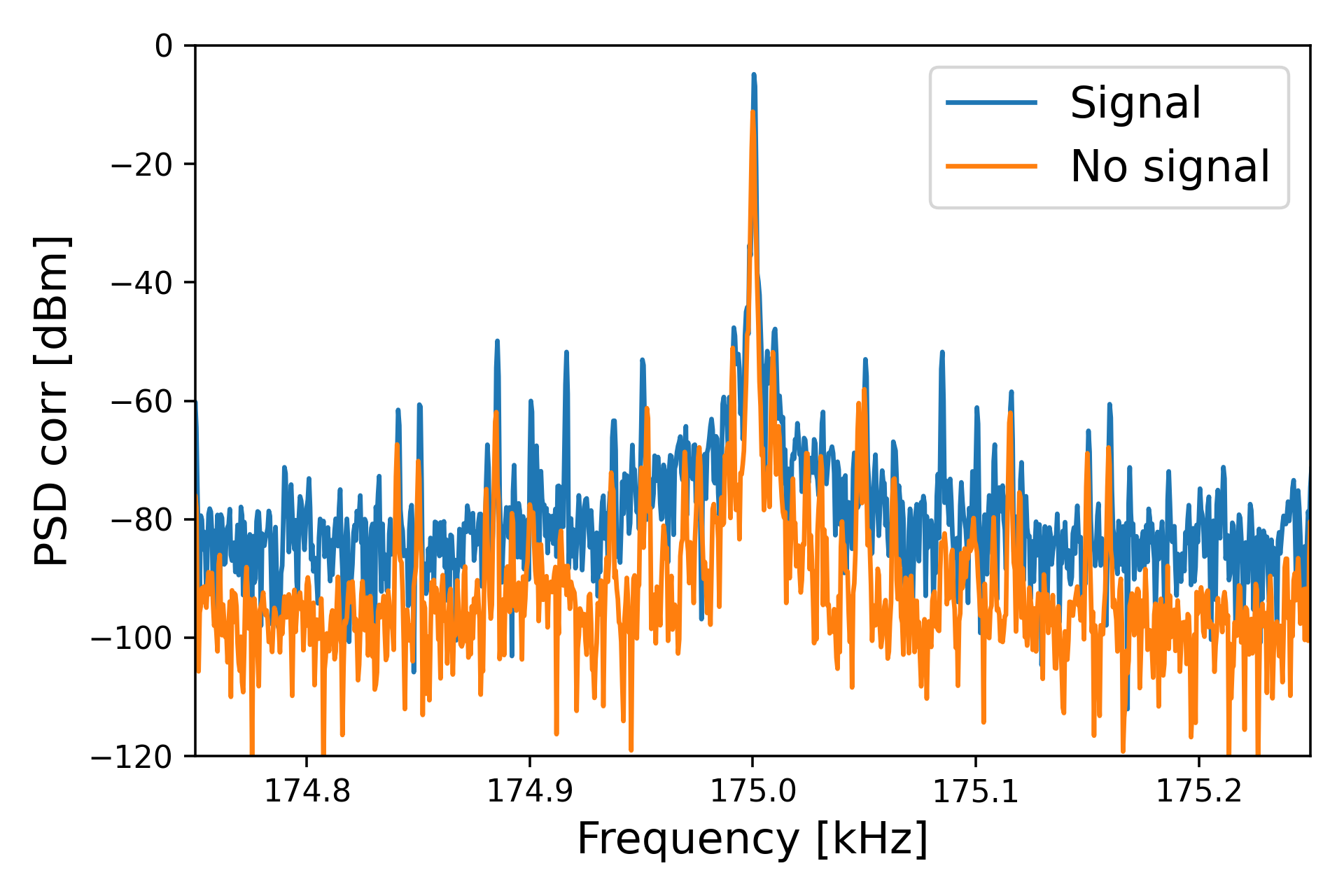}
	\end{minipage}
	\hfill
	\begin{minipage}{0.49\textwidth}
		\includegraphics[width=1\linewidth]{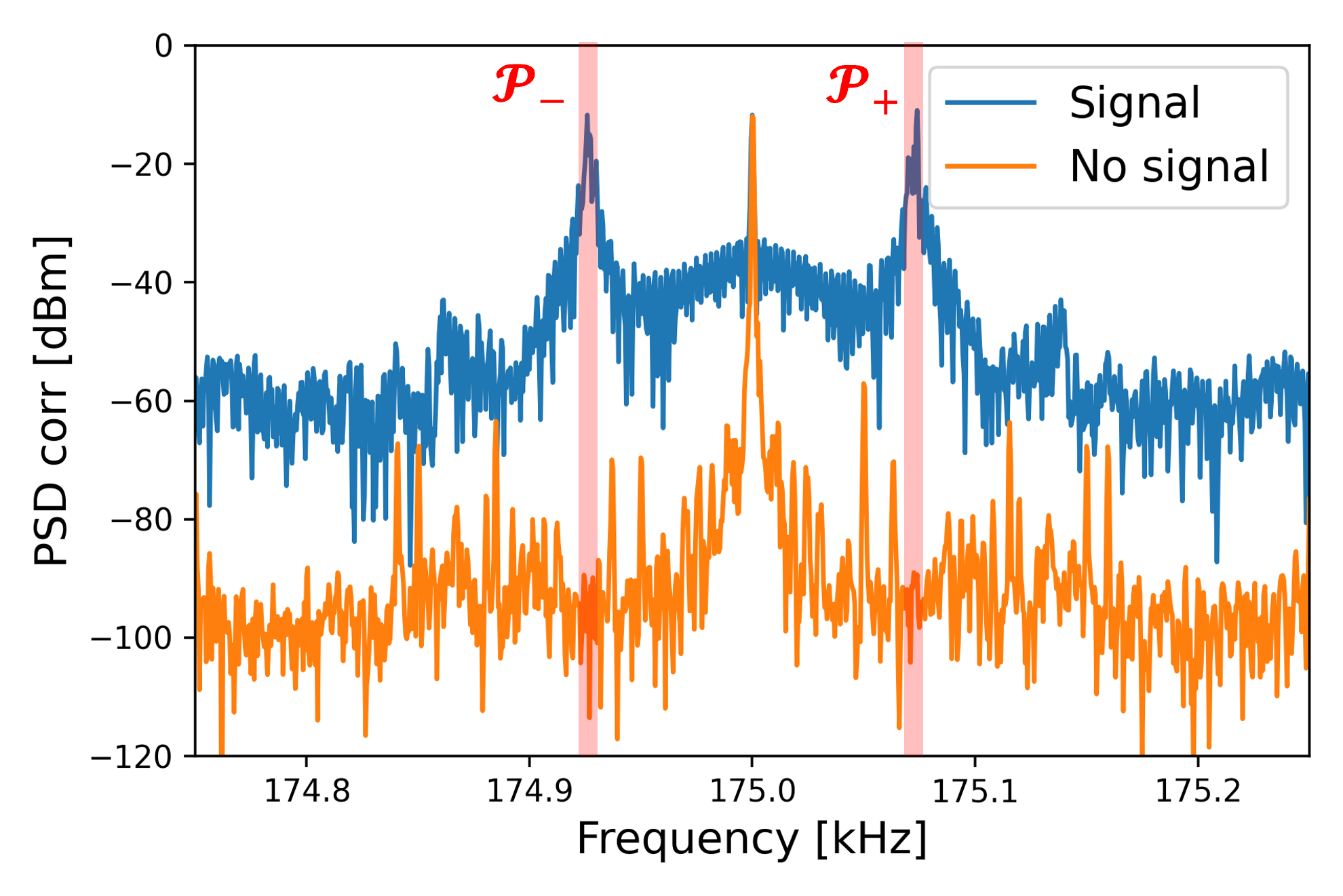}
	\end{minipage}
	\caption{Examples of average PSDs at the output of the correlator. The science source is QCL B in laser regime. The integration time is $1~\text{s}$ to obtain $1~\text{Hz}$ frequency resolution. \emph{(Left)} Without any path-length modulation of the local oscillator. \emph{(Right)} $\text{PSD}_{\text{corr}}$ With path-length modulation of the local oscillator at $\pm70~\text{Hz}$ with a dither mirror.}
	\label{fig:correlation_peak}
\end{figure*}

We read the correlation information from the output of the balanced photodetector. In the absence of a local oscillator path-length modulation, the correlation signal is encoded at $80.175~\text{MHz}-80.000~\text{MHz}=175~\text{kHz}$, as shown on Fig.~\ref{fig:correlation_peak} \emph{(Left)}. When the $70~\text{Hz}$ local oscillator path-length modulation is active, the correlation signal is encoded at $f_-=174.930~\text{kHz}$ and $f_+=175.070~\text{kHz}$. In our current set-up, the path-length modulation is far from perfectly linear because of the acceleration and deceleration phase of the dither motor and because of vibrations: the correlation signal is spread across a wider range of frequencies, as shown on  Fig.~\ref{fig:correlation_peak} \emph{(Right)}. 

\subsection{Correlator data acquisition}

Data acquisition from the correlator is performed with a digital oscilloscope. The correlation signal $i_{\text{BPD}}(t)$ is recorded in $200~\text{ms}$ series with $625~\text{kHz}$ sampling rate. The PSD is computed, leading to a spectrum with $5~\text{Hz}$ resolution. The power spectra of the series are averaged, resulting to the incoherent integration of the correlation signal. The measurement noise is estimated by computing the standard deviation $\sigma$ of the different power spectra series. When performing $N$ measurements, the uncertainty on the measured average value will be given by $\sigma/\sqrt{N}$.

\section{Results}
\label{section:results}

In the following, we describe the characterization steps that validate our proof
of concept set-up to correlate the mid-infrared signals from two telescopes.

\subsection{Photonic correlator response measurement}

The response of the photonic correlator was measured separately from the rest of the system to study its performance. To do so, we use RF signal generators as inputs for the correlator. In practice, we measure separately the responses to correlated and uncorrelated inputs respectively corresponding to the previously mentioned $f_{\text{corr}}$ and $f_{\text{noise}}$ functions. Similarly, we obtain the $f_{\text{noise}}$ function by generating uncorrelated white noises $w_1$ and $w_2$ at the inputs.

\subsubsection{Linear response to correlated signals}

\begin{figure*}
	\centering
	\begin{minipage}{0.49\hsize}
		\includegraphics[width=1\linewidth]{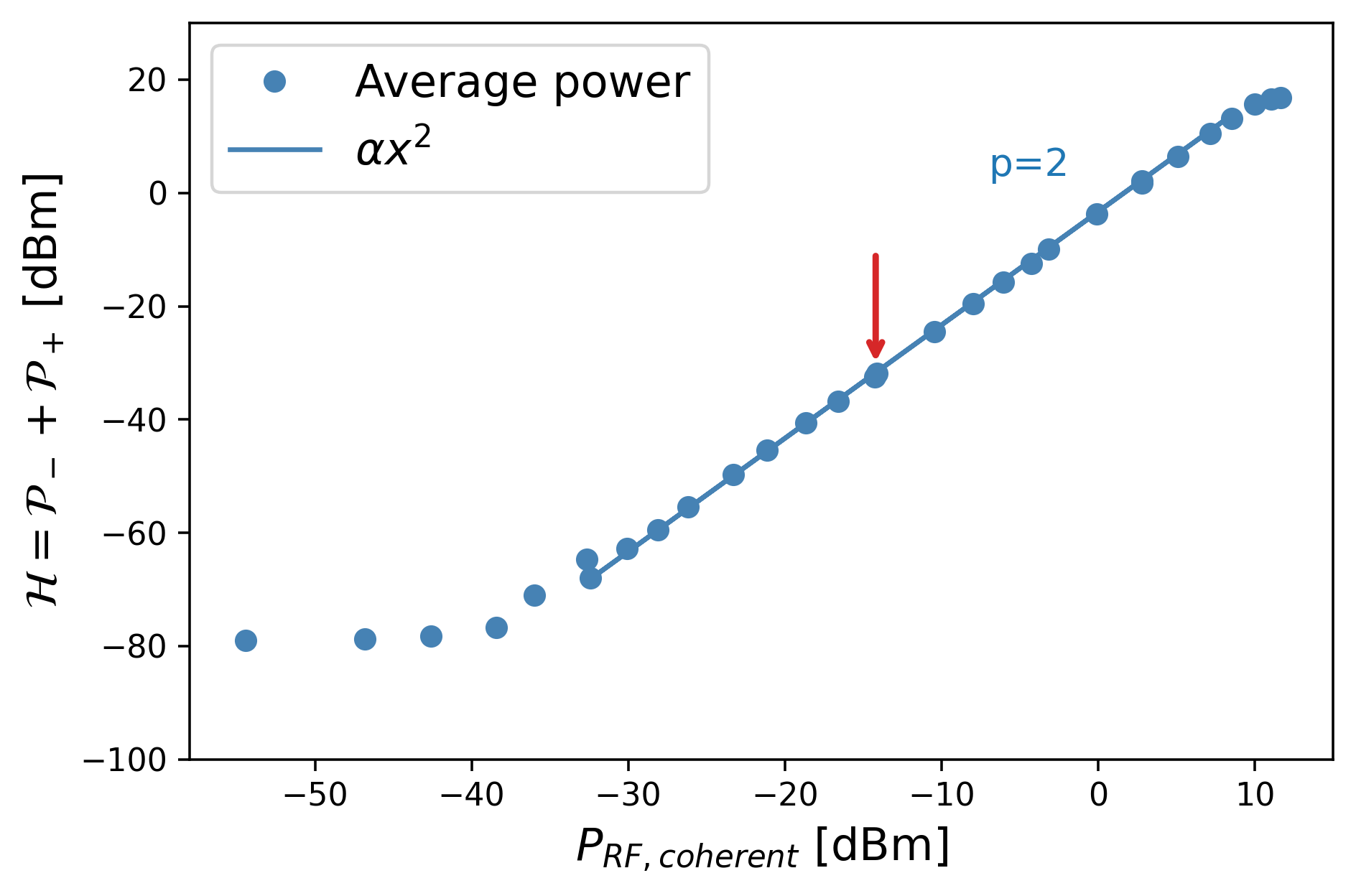}
	\end{minipage}
	\hfill
	\begin{minipage}{0.49\textwidth}
		\includegraphics[width=1\linewidth]{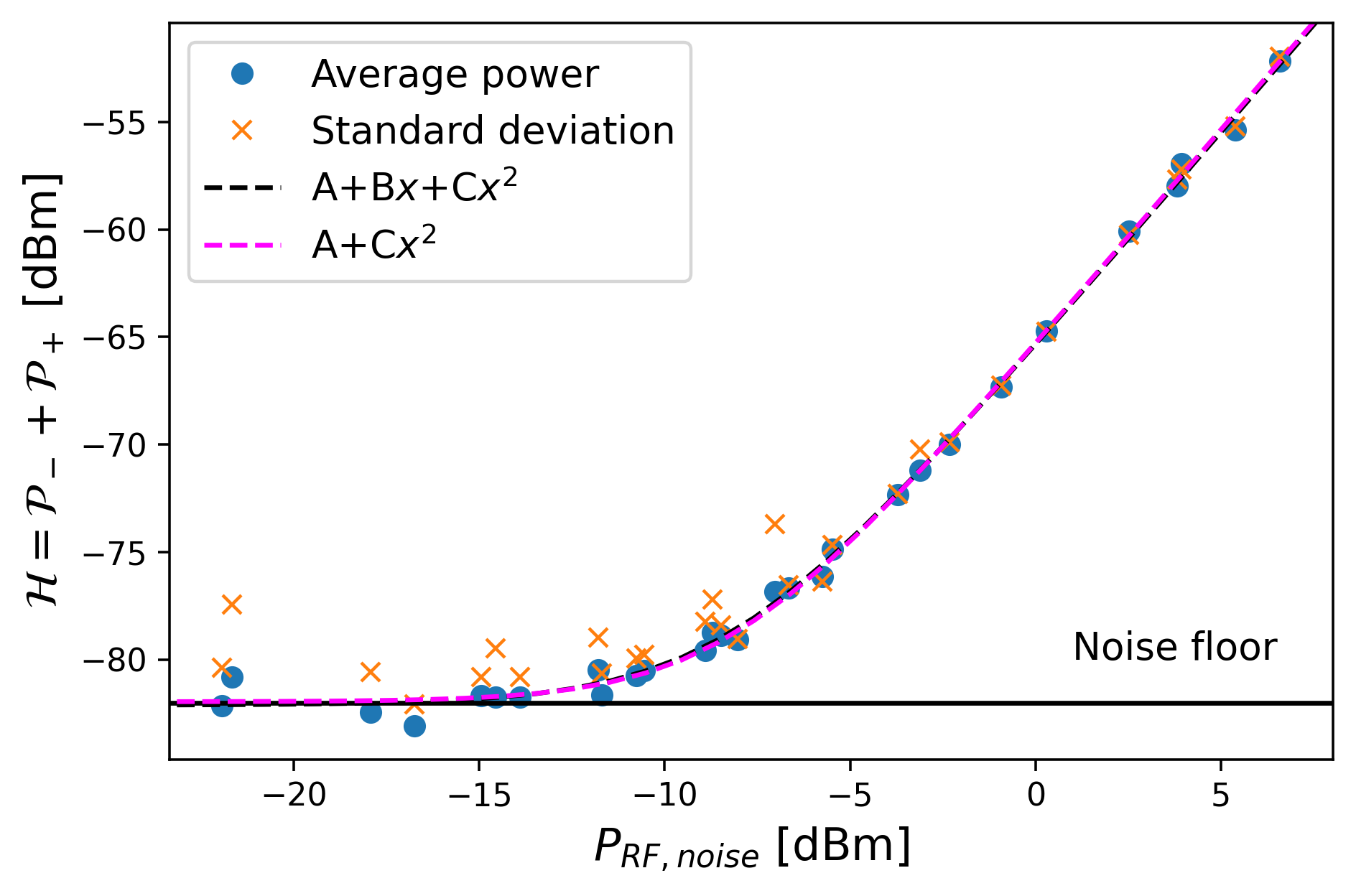}
	\end{minipage}
	\caption{\emph{(Left)} Experimental measurement of $\mathcal{H}=\mathcal{P}_-+\mathcal{P}_+$ as a function of the RF coherent power $P_{\text{RF,coherent}}$. Two sinewaves representing $h_1$ and $h_2$ were used as fully correlated inputs such that $P_{\text{RF,coherent}}=\sqrt{P_{\text{RF},1}P_{\text{RF},2}}$. \emph{(Right)} Experimental measurement of $f_{\text{noise}}$ as a function of the uncorrelated input RF power $P_{\text{RF,noise}}$. Two independent white noises (representing $w_1$ and $w_2$) were used as uncorrelated inputs such that $P_{\text{RF,noise}}=\sqrt{P_{\text{RF},1}P_{\text{RF},2}}$.}
	\label{fig:correlation_nep}
\end{figure*}

The response of the correlator to coherent signals was measured using two RF sine functions $h_1$ and $h_2$ at $119~\text{MHz}$ with a $70~\text{Hz}$ frequency difference. The frequency difference emulates the path-length modulation from the dither mirror. By measuring the correlator output as a function of coherent RF power $P_{\text{RF,coherent}}=\sqrt{P_{\text{RF,}1}P_{\text{RF,}2}}$, we obtain the $f_{\text{corr}}$ function as shown on Fig. \ref{fig:correlation_nep} \emph{(Left)}. We can decompose the resulting $f_{\text{corr}}$ function into three main regime: 
\begin{itemize}
    \item \textbf{A linear regime from $-40$ to $10~\text{dBm}$ of coherent input power.} In this regime $f_{\text{corr}}$ scales, as expected, proportionally to ${P_{\text{RF,coherent}}}^2$ ($p=2$ slope in log-scale). \textbf{This five order of magnitude linear range shows the photonic correlator is capable of handling RF signals with a wide range of powers.} In addition, strong imbalance between $P_{\text{RF},1}$ and $P_{\text{RF},2}$, with power ratio $P_{\text{RF},1}/P_{\text{RF},2}$ up to $10^5$ (the point indicated by the red arrow on Fig. \ref{fig:correlation_nep} \emph{(Left)}), did not lead to a deviation from the linear curve. For an operational point of view, this means \textbf{it is not necessary to balance the RF signals or the optical flux on the mid-infrared detectors}.
    
    \item \textbf{A saturation regime above $10~\text{dBm}$ of coherent input power.} The saturation comes from the balanced photodetector, and from the amplitude modulators because the RF signal amplitude becomes comparable to the $V_{\pi,k}$ values.
    
    \item \textbf{A noise floor regime below $-40~\text{dBm}$ of RF input power} where the $-80~\text{dBm}$ noise of the photonic correlator dominates the correlation signal.
\end{itemize}

\subsubsection{Photonic correlation noise robustness}

We used RF signal generators to generate two $w_1$ and $w_2$ uncorrelated white noise signals at the RF inputs of the photonic correlator to retrieve the $f_{\text{noise}}$ function from equation (\ref{eq:correlator_calibration}). As shown on Fig.~\ref{fig:correlation_nep} \emph{(Right)}, we measured the average power of $P_{\text{corr}}$ as a function of uncorrelated white noise RF input power.
White noise RF input powers lower than $-10~\text{dBm}$ do not affect the $\simeq-80~\text{dBm}$ noise floor of the correlator. At RF power that are higher than $-5~\text{dBm}$, the noise floor scales with a $p\simeq 2$ slope, as expected from equation (\ref{eq:f_noise}). The curve was fitted with $A+Cx^2$ and $A+Bx+Cx^2$ polynomials. The $B$ term is negligible compared with the other terms. Thus, we can consider $B=0$. This shows that, \textbf{as long as the RF coherent input power is sufficiently high for the noise floor to be negligible, the correlation operation is performed at the theoretical limit}. We obtain $A=(6.4\pm0.2)\times10^{-11}~\text{W}$ and $C/\Delta\nu_{\text{noise}}=(2.9\pm0.1)\times10^{-4}~\text{W}^{-1}$.

The $1/\Delta\nu_{\text{noise}}$ dependence was verified with a numerical simulation, as described in Appendix~\ref{app:simulation}. The numerical model considered noise-less Mach-Zehnder modulators with the measured extinction ratio of our real modulator, noise-less AOFSs and noise-less balanced photodetector. White-noise inputs were used as RF inputs. 

\begin{figure*}
	\centering
	\begin{minipage}{0.49\hsize}
		\includegraphics[width=1\linewidth]{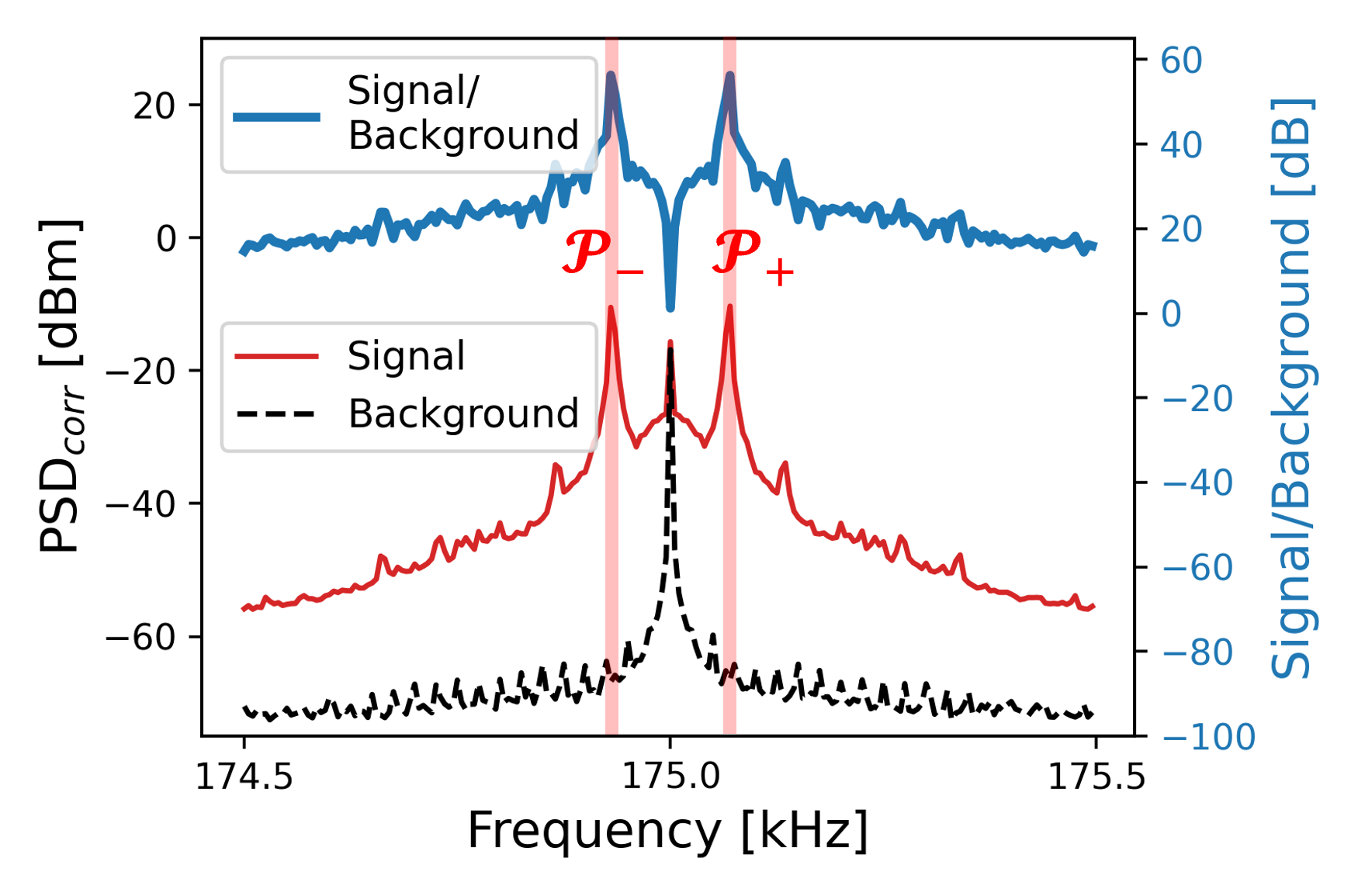}
	\end{minipage}
	\hfill
	\begin{minipage}{0.49\textwidth}
		\includegraphics[width=1\linewidth]{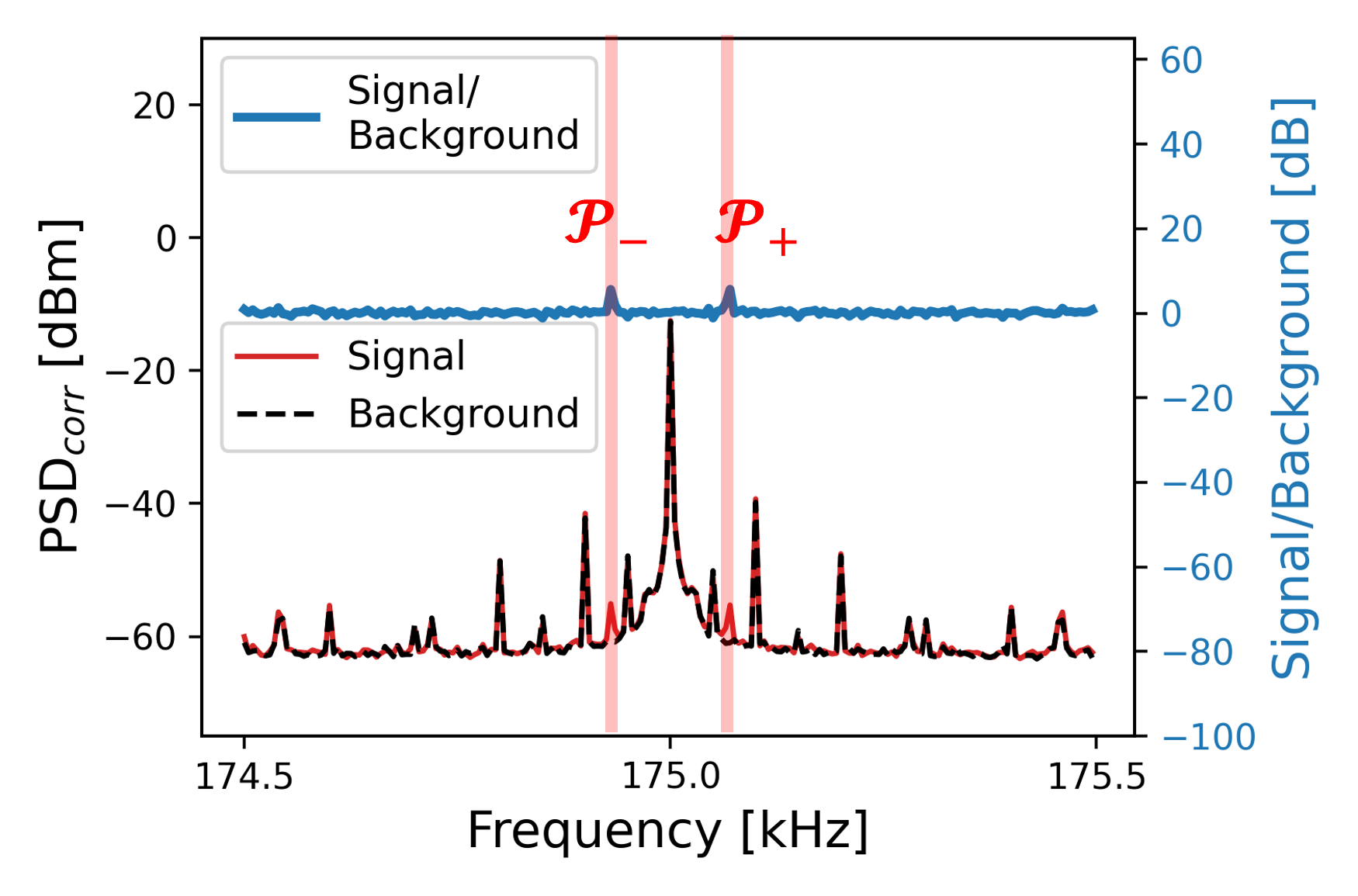}
	\end{minipage}
	\caption{Average PSDs of the correlator output ($\text{PSD}_{\text{corr}}$) in the presence of science signal and in its absence (background). The ratio between the output in the presence of the signal and the background is also represented: \emph{(Left)} PSDs for the verification of the correlator calibration with high coherent RF signal. \emph{(Right)} PSDs for the detection limit measurement with low coherent RF signal.} 
	\label{fig:correlator_output}
\end{figure*}

\subsubsection{Correlator Noise Equivalent Power and Noise Equivalent Input}

We can define the Noise Equivalent Power of the correlator $\text{NEP}_{\text{corr}}$ in the form of equation (\ref{eq:nep_corr}). It corresponds to the correlated RF power $P_{\text{RF,coherent}}$ that would lead to having an SNR of 1 at the output of the correlation in the case where $w_1=w_2=0$.

\begin{equation}
    \text{NEP}_{\text{corr}}=\frac{1}{\sqrt{G_1G_2}}\sqrt{\frac{A}{2\alpha_{\text{corr}}}}
    \label{eq:nep_corr}
\end{equation}

\noindent Assuming $G_1=G_2=3.1\times 10^4=45~\text{dB}$, we obtain $\text{NEP}_{\text{corr}}=-88~\text{dBm}=2\times 10^{-12}~\text{W}$.

Similarly, we can define the Noise Equivalent Input of the correlator $\text{NEI}_{\text{corr}}$ as equation (\ref{eq:nei_corr}). $\text{NEI}_{\text{corr}}$ corresponds to the RF power of the noise that would be equivalent to the noise floor of the correlator. This metric enables us to compare the correlator noise to other noise levels in the system such as detection noise and amplification noise.

\begin{equation}
    \text{NEI}_{\text{corr}}=\frac{1}{\sqrt{G_1G_2}}\sqrt{\frac{A\Delta\nu_{\text{noise}}}{C}}
    \label{eq:nei_corr}
\end{equation}

\noindent  Assuming $G_1=G_2=3.1\times 10^4=45~\text{dB}$, we obtain $\text{NEI}_{\text{corr}}=-53~\text{dBm}=5\times 10^{-9}~\text{W}$.

It should be noted that, if we could operate the system at a higher modulation frequency $f_{12}$, away from the central parasitic peak at $175~\text{kHz}$, the noise equivalent input would be decreased. For example at $f_{12}= 7~\text{kHz}$, which corresponds to maximal fringe frequency for a kilometric baseline, would decrease $\text{NEI}_{\text{corr}}$ by nearly $4~\text{dB}$.

\subsection{System calibration}

Calibration of the overall instrument is primordial to retrieve the optical
coherent flux $F_c$ directly from the observable $H$ at the output of the correlator. To do so, we place ourselves in a situation where the heterodyne signals are perfectly correlated (the degree of coherence $\gamma_{12}$ is equal to $1$). This is achieved by making sure the interferometric delay between the two detectors is shorter than the coherence length of the science source. This can typically be done by setting the science QCL in laser regime. By measuring the optical powers impinging on the detectors and the correlation power $H$ at the output of the correlator, we can obtain the value of $\alpha_{\text{IR}}\alpha_{\text{corr}}$ in equation (\ref{eq:f_corr}) linking $H$ to $F_c$.

Alternatively, one may measure $\alpha_{\text{IR}}$ and $\alpha_{\text{corr}}$ separately by measuring the response of the different elements of the system though it makes the process more tedious.

The calibration procedure typically needs to be done every time the alignment of the mid-infrared stage is changed. Notably, the calibration must be completed when the delay line is adjusted because of the sensitivity of the system to any mismatch between the local oscillator and the science signal. For recurrent measurements where the alignment is changed, one may simply measure the variation $\alpha_{\text{IR}}$ with the alignment instead of measuring the whole $\alpha_{\text{IR}}\alpha_{\text{corr}}$ product.

We obtain $\alpha_{\text{corr}}=1190\pm70~\text{W}^{-1}$ with a laser-laser heterodyne signal at $300~\text{MHz}$ and $\alpha_{\text{IR}}=3.0\pm0.1\times10^{14}$. Details of this measurement are described in Appendix~\ref{app:correlator_laser_callibration}.

\subsection{ASE coherence envelop and fiber delay compensation}

To prove that the system does measure the mid-infrared coherent flux, we measured the degree of coherence $\gamma_{12}(\Delta\tau)$ between the signals from detectors 1 and 2 at different free-space delays $\Delta\tau_{\text{FS}}$, expecting the recover the coherence envelop of the ASE source of known optical bandwidth. 

The degree of coherence $\gamma_{12}(\Delta\tau)$ is obtained using equation (\ref{eq:degree_of_coherence}). $P_{\text{RF,coherent}}$ is calculated based on the correlator output while $P_{\text{RF,science},1}$ and $P_{\text{RF,science},2}$ are directly measured with an oscilloscope.

\begin{equation}
    \gamma_{12}=\frac{P_{\text{RF,coherent}}}{\sqrt{P_{\text{RF,science},1}P_{\text{RF,science},2}}}
    \label{eq:degree_of_coherence}
\end{equation}

The free-space delays are adjusted up to $95~\text{cm}$ with the free-space optical delay line acting on $\Delta\tau_{\text{FS}}$. Two sets of measurements were carried out, one of them with an extra fiber of length $23.5~cm$ in one arm of the correlator, between the AOFS and the fiber coupler on channel 1, to test the possibility to compensate mid-infrared free-space delay with fibre delay in the photonic correlator. We expected to obtain identical coherence envelopes for the two sets on measurements, expect for an offset corresponding to the added fibre delay.

\begin{figure}
	\centering
    \includegraphics[width=0.7\linewidth]{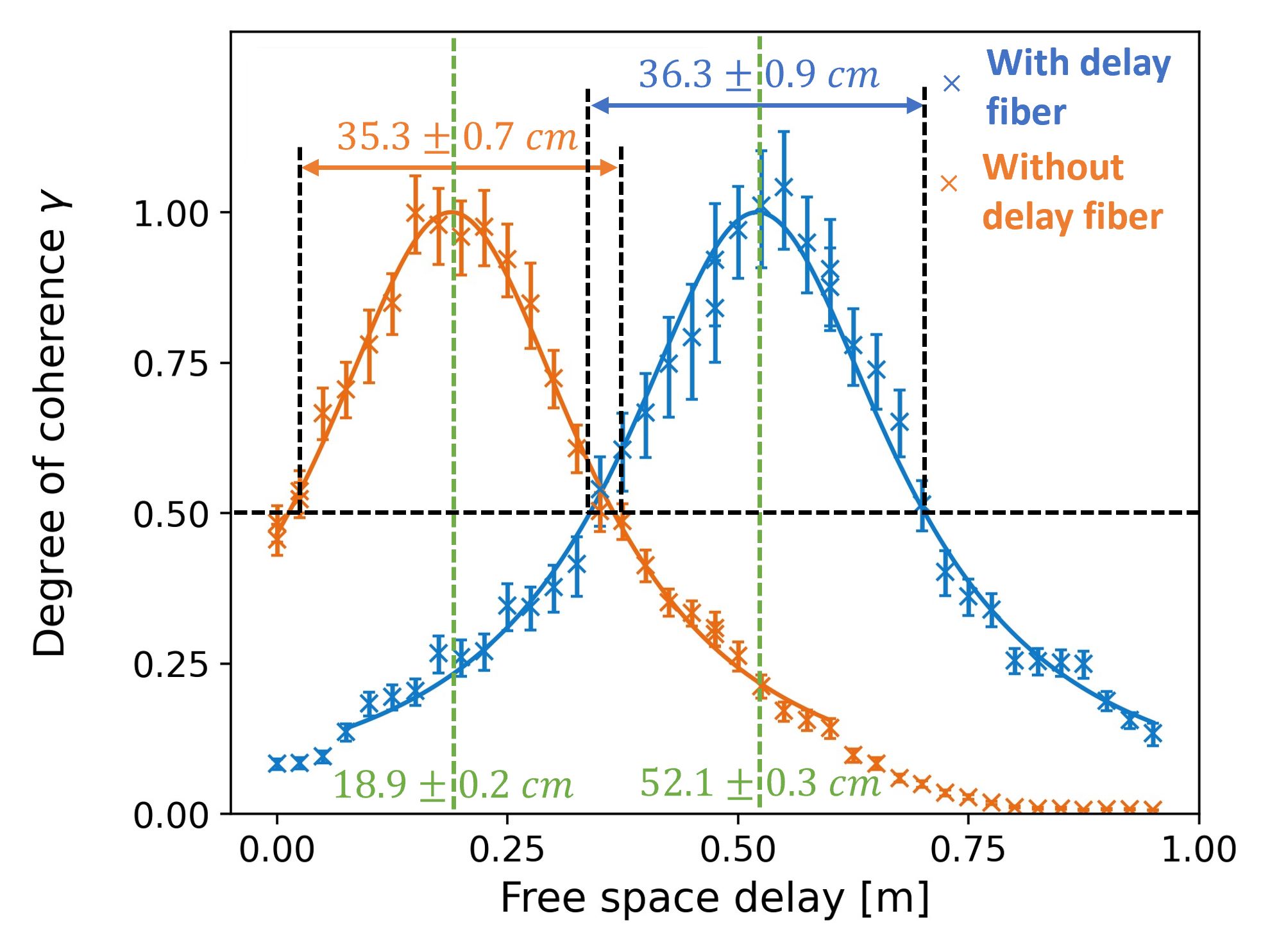}
    \caption{Measurement of the coherence envelop of ASE radiation in the presence or absence of a $23.5~cm$ delay fiber in the photonic correlator.}
	\label{fig:coherence_envelop}
\end{figure}

The results of these measurement are shown in Fig.~\ref{fig:coherence_envelop}. The data curves were fitted with Lorentzian functions using the least mean square method. We obtained a Full Width Half Maxima (FWHM) of $36.3\pm 0.9~cm$ with the delay fiber and $35.5\pm 0.7~cm$ without it. These FWHM values are compatible with the estimated $300~\text{MHz}$ optical bandwidth of the ASE signal for a QCL B current of $902.0~\text{mA}$. 

The resemblance between the two curves indicates that in the presence of an extra delay $\Delta\tau_{\text{corr}}$ in the correlator, we can recover the same coherence state as without that delay applying an free-space delay $\Delta\tau_{\text{FS}}$ and vice-versa. The offset between the two curves on Fig.~\ref{fig:coherence_envelop} corresponds to the free space delay that should be applied. Thus, \textbf{interferometric delay compensation in heterodyne interferometry can be achieved with fiber delays lines in the photonic correlator}.

\subsection{Optical detection limit}

The overall detection limit of the system was estimated by decreasing the optical power of the ASE radiation using optical densities until no correlation peak could be observed at the output of the correlator. By precise calibration of the system, we can measure the minimum measurable correlation peak and recover the corresponding coherent flux $F_c$ which we consider is the detection limit of the system.

We minimized the delay $\Delta\tau$ to place the set-up in a $\gamma_{12}\simeq1$ situation. The intermediate frequency RF signals $h_1$ and $h_2$ were amplified by $45~\text{dB}$. We acquired $250$ "sequences" of $200~\text{ms}$ integration time of signal at the output of the correlator. The resulting average PSD at the output of the correlator is shown on Fig. \ref{fig:correlator_output} \emph{(Right)}. Details of the measurements are described in Appendix \ref{app:detection_limit}. We measure an average value of $H=(6.65\pm0.45)\times 10^{-9}~\text{W}$%=-51.8\pm0.3~\text{dBm}$
with a standard deviation of $6.1\times 10^{-9}~\text{W}$. This corresponds to a SNR of $1.25$ on the estimation of $H$ and to an SNR of $2.5$ on the estimation of $F_c$. Thus, this measurement is not exactly at the detection limit of the system (usually considered to be SNR$=1$), but it is the lowest signal we were able to detect. 

We obtain $P_{\text{RF,coherent}}=(7.75\pm0.35)\times 10^{-11}~\text{W}$ (before amplification). Using the calibration of the infrared heterodyne detection stage, we calculate the corresponding coherent flux $F_c$ and \textbf{we obtain $\mathbf{F_c=140\pm 30}~\text{fW}$ which we will consider to be the detection limit of our system for an integration time of $\mathbf{200}~\text{ms}$}. For an astronomical object, with the $900~\text{MHz}$ bandwidth of our detectors and a $8~\text{m}$ telescope with $10~\%$ transmission (from pupil to detector), this would correspond to a flux of approximately $1800~\text{Jy}$.

\section{Discussion}
\label{section:discussion}

\begin{figure}
	\centering
	\includegraphics[width=0.7\linewidth]{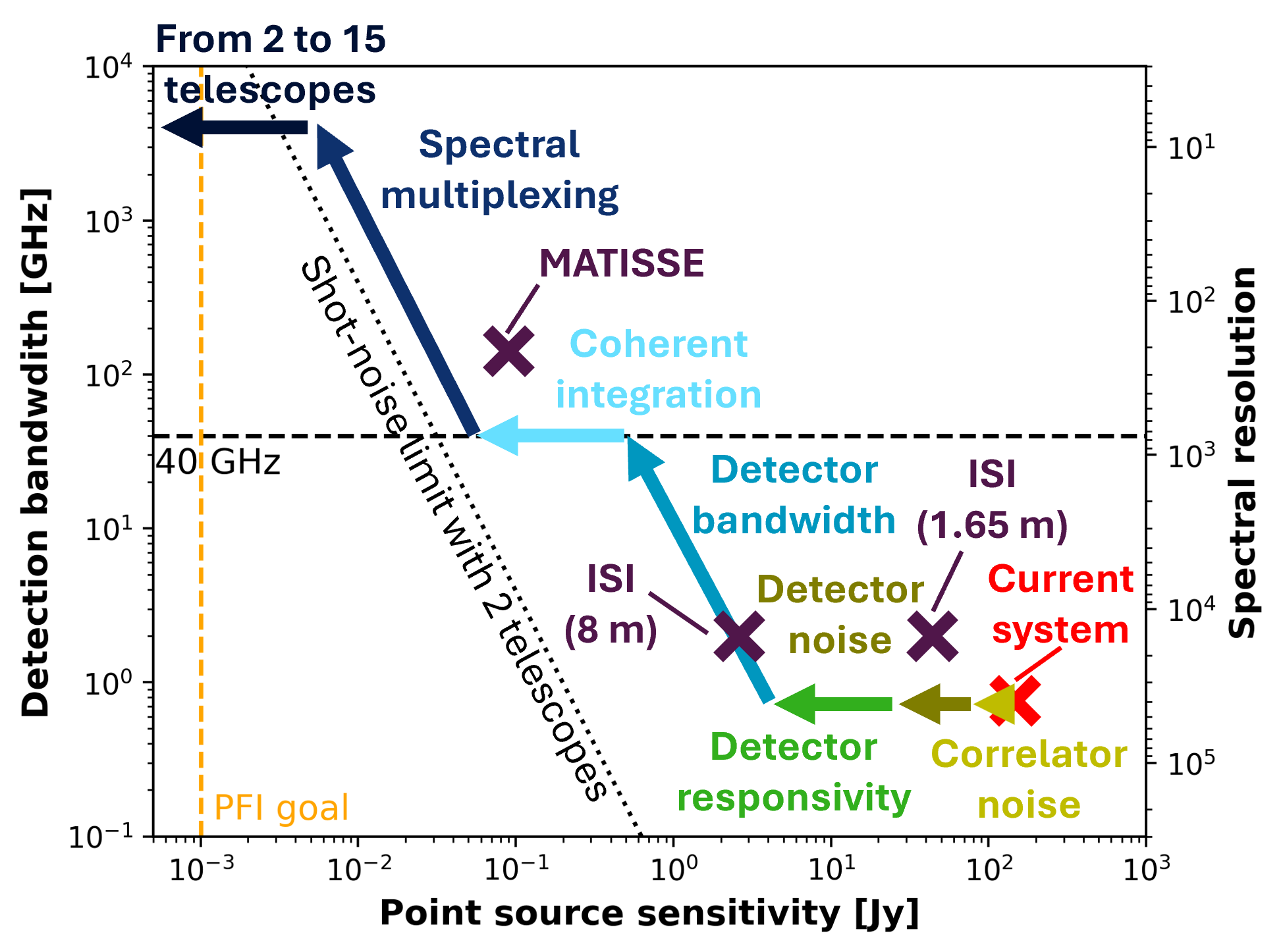}
	\caption{Estimated point source sensitivity of our current system if it were used with 8 m class telescopes with 10\% transmission. The impact of different key technological improvements for the advent of mid-infrared interferometry is displayed.
	}
	\label{fig:roadmap}
\end{figure}

Our experimental results have shown that, from a functional point of view, it is
possible to correlate mid-infrared electromagnetic signals coming from two
separated apertures using a photonic correlator. While we have not done the
exercise we are confident that this correlation could be done over a kilometric
baseline. We discuss now what limits the performance and how such limits can be
overcome in the context of an actual astronomical instrument. The roadmap for the development of an actual PFI-like heterodyne interferometer is illustrated by Fig. \ref{fig:roadmap}. The calculated detection limit is based on the use of 8 m class telescopes for an integration time of one-hour. The role of key-technologies is highlighted to lower our current detection limit from $130~\text{Jy}$ to less than $1~\text{Jy}$. The detection limits of ISI and MATISSE instruments are also represented for comparison. In the case of ISI, which used 1.65 m telescopess, we also extrapolated the detection limit for 8 m telescopes for fair comparison.

\subsection{Infrared detection limitations}

\begin{table*}[]
	\centering
	\footnotesize
	\begin{tabular}{|l||c|c||c|cc|c||c|c||c|}
		\hline
		\multicolumn{1}{|c||}{Detector} 
		& \multicolumn{1}{c|}{$\eta$} 
		& \multicolumn{1}{c||}{$\Delta\nu_{\mathrm{det}}$} 
		& \multicolumn{1}{c|}{$T_{\mathrm{det}}$} 
		& \multicolumn{1}{c|}{$T_{\text{amp}}$} 
		& \multicolumn{1}{c|}{$T_{\text{shot-noise}}$} 
		& \multicolumn{1}{c||}{$n_{F,\text{det}}$}
		& \multicolumn{1}{c|}{$T_{\text{corr}}$}
		& \multicolumn{1}{c||}{$n_{F,\text{corr}}$}
		& \multicolumn{1}{c|}{Point source}
		\\
		& \multicolumn{1}{c|}{$[\text{\%}]$}
		& \multicolumn{1}{c||}{$[\text{GHz}]$}
		& \multicolumn{1}{c|}{$[\text{K}]$}
		& \multicolumn{1}{c|}{$[\text{K}]$} 
		& \multicolumn{1}{c|}{$[\text{K}]$}
		& \multicolumn{1}{c||}{}
		& \multicolumn{1}{c|}{$[\text{K}]$}
		& \multicolumn{1}{c||}{}
		& \multicolumn{1}{c|}{sensitivity $[\text{Jy}]$}
		\\ \hline\hline
		VIGO PVI-4TE-10.6
		& $6.5\%$
		& $0.9$  
		& \multicolumn{2}{c|}{$350$}       
		& $160$
		& $3.2$    
		& $20$
		& $1.0$
		& $130$
		\\ \hline
		QWIP (\cite{Hakl2021}) 
		& $7\%$    
		& $30$
		& $5000$                      
		& \multicolumn{1}{c|}{$1000$}   
		& $2200$ 
		& $3.7$ 
		& $3700$
		& $1.5$
		& $35$
		\\ \hline
		HgCdTe (KV-104-0.1-1E/11)  
		& $50\%$  
		& $0.5$      
		& $\sim78$                     
		& \multicolumn{1}{c|}{$50$}    
		& $5600$
		& $1.0$  
		& $400$
		& $1.1$
		& $15$
		\\ \hline
		Ideal detector ($10~\text{GHz}$)    
		& $100\%$ 
		& $10$      
		& $0$                         
		& \multicolumn{1}{c|}{$1000$}
		& $25000$
		& $1.0$
		& $8000$
		& $1.3$
		& $2$
		\\ \hline
		Ideal detector ($100~\text{GHz}$)   
		& $100\%$ 
		& $100$                      
		& $0$                         
		& \multicolumn{1}{c|}{$1000$}
		& $25000$
		& $1.0$
		& $25000$
		& $2.0$
		& $0.5$
		\\ \hline
	\end{tabular}
	\normalsize
	\caption{Detectors used for sensitivty comparison. Point source sensitivities were computed considering a two telescope heterodyne interferometer with 8 m telescopes and one hour integration time.}
	\label{tab:detector_parameters}
\end{table*}

The performance of our detectors are compared with other detectors in Tab.\ref{tab:detector_parameters}. $T_{\mathrm{det}}$ is the noise temperature of the detector; $T_{\text{amp}}$ is the equivalent noise temperature of the RF amplifiers; $T_{\text{shot-noise}}$ is the equivalent noise temperature of the shot-noise on the detector; $T_{\text{corr}}$ and $n_{F,\text{corr}}$ are respectively the equivalent noise temperature and noise factor of the photonic correlator which will be discussed in the following section. We define the noise factor of the detector $n_{F,\text{det}}$ as the ratio between the sum of all noise sources related to detection and the shot-noise. In an ideal detection scheme, shot-noise is the main source of noise ($T_{\text{shot-noise}}\gg T_{\mathrm{det}}+T_{\text{amp}}$) such that $n_{F,\text{det}}\simeq1$.

We calculated the expected point source sensitivity for a two telescope system considering $8~\mathrm{m}$ telescopes with $30~\%$ transmission and a $3600~\text{s}$ integration time (incoherent integration of $18000$ sequences of $0.2~\text{s}$). The RF gains $G_1$ and $G_2$ were set to maximize the correlator RF input power without entering saturation regime.

We selected two detectors: the \textit{KV-104-0.1-1E/11} commercial HgCdTe detector from \textit{Kolmar Technologies} with $\Delta\nu=500~\text{MHz}$ bandwidth and the state-of-the-art Quantum Well Infrared Photo-detector (QWIP) that was described in \cite{Hakl2021} whose bandwidth reaches $\Delta\nu=30~\text{GHz}$ at $78~\mathrm{K}$. We also compared these detectors with an ideal noiseless detector with $100~\%$ QE.

We are particularly interested in QWIPs detectors for their capacity to detect large bandwidth signals, up to more than $100~\text{GHz}$ as reported by \cite{Lin2023,Grant2006}. QWIPs rely on high speed intersubband processes whereas usual photodiodes utilize electronic transitions between the valence and the conduction band (\cite{Schneider2007}). QWIPs are still an active field of research, with on-going works to improve their responsivity and bandwidths
(\cite{Lin2023,Palaferri2018,Rodriguez2022,Hakl2021}). Recently, the use of
meta-material engineering was proven effective to improve the responsivity of
QWIP detectors by coupling the electrical field of the incoming signal into the
active materials of the detector. Additionally, the use of meta-material
engineering reduces the electrical surface of the detector and thus increases
its electrical bandwidth by reducing its capacitance (\cite{Rodriguez2022}).

The current main limitation of our system comes from the $\eta\simeq6.5\%$ low quantum efficiency of our detectors: it reduces the signal power and it prevents reaching the shot-noise limited regime ($n_{F,\text{det}}=3.2$), especially because of the low $1~\text{mW}$ saturation threshold of the detector. QWIPs also suffer from their low quantum efficiencies ($\eta\lesssim 10\%$) and high noise levels ($n_F\simeq 4$). They prevent them from reaching the shot-noise limited regime, but their large bandwidths up to $100~\text{GHz}$ (\cite{Lin2023}) enables them to detect a larger chunk of the wide-band astronomical signal. Only the HgCdTe detectors like the \textit{KV-104-0.1-1E/11} reach a regime where both correlator and detector noise sources are negligible compared with shot-noise ($n_{F,\text{det}}=1.0$).

The impact of implementing new types of detectors on our demonstration bench was studied using the previously measured photonic correlator response.  
The detection limits are approximately $130~\text{Jy}$ for the current detector with $900~\text{MHz}$ bandwidth, $15~\text{Jy}$ with the commercial HgCdTe detector with $500~\text{MHz}$ bandwidth and $35~\text{Jy}$ with the QWIP at
$30~\text{GHz}$. In comparison, an ideal detector would have a detection limit of $2~\text{Jy}$ at $10~\text{GHz}$ (corresponding to R=3000 resolution) and $0.5~\text{Jy}$ at $100~\text{GHz}$ (corresponding to R=300 resolution). Therefore, implementation of high-end QWIPs or HgCdTe detectors could improve the performance of the system by an order of magnitude, with another order of magnitude that could be gained by improving the quantum efficiency of QWIPs, notably using meta-materials (\cite{Rodriguez2022}). It should be noted that the indicated detection limits remain

Another limitation of our current IR detection stage is the polarization selection of the astronomical signal caused by the polarization of the LO. A straight-forward solution would consist in using a polarization beam splitter and a detection and correlation set-up for each polarization.

\subsection{Photonic correlator}

The photonic correlator is able to perform the correlation of $10~\text{GHz}$ signals. Similar Mach-Zehnder intensity modulators with bandwidths up to 40 GHz are commercially available. Some with more than $100~\text{GHz}$ have even been demonstrated by \cite{Valdez2022} and could be utilized if high-bandwidth detectors such as QWIPs are used.

The photonic correlator, operating at $10~\text{dBm}=10~\text{mW}$ RF input power has an approximate noise rejection of $2000$ which means that the correlator can detect a signal 2000 times fainter than the noise level.

The correlator Noise Equivalent Input can be converted to an equivalent noise temperature $T_{\text{corr}}$ at the level of the IR detectors. We then define the noise factor of the correlator $n_{F,\text{corr}}$ as the ratio between all noises in the system and all noises except correlator. Having $n_{F,\text{corr}}=2$ means the photonic correlator degrades the SNR by a factor of two compared with an ideal correlation operation. A higher quantum efficiency for the detector leads to a higher heterodyne signal and lower RF amplification. Thus, the equivalent noise temperature of the correlator increases since it scales inversely to the RF gain. This does not degrade the noise factor because the shot-noise equivalent temperature also scale linearly with the quantum efficiency. However, a larger detection bandwidth does degrade $n_{F,\text{corr}}$ because, for a given RF power at the input of the correlator, the larger the signal bandwidth, the lower the noise at the output of the correlator.

For a fair comparison, we consider that we can operate our system at $f_{12}=7~\text{kHz}$ rather than our current $f_{12}=70~\text{Hz}$ (limited by the dither mirror). For the current state of the system, we obtain $n_{F,\text{corr}}=1.0$, meaning the photonic correlator does not degrade the performance of the system. In the case of the QWIP from \cite{Hakl2021} with $G=10^6=60~\text{dB}$ and a bandwidth of $30~\text{GHz}$, we obtain a equivalent temperature $T_{\text{corr}}=3700~\text{K}$ which is higher than the shot-noise equivalent temperature $T_{\text{shot-noise}}=2200~\text{K}$ but lower than the detector noise. In the end, this leads to $n_{F,\text{corr}}=1.5$ meaning the photonic correlator would have a slight impact on the SNR. With an ideal detector of $100~\text{GHZ}$ bandwidth, the correlator noise is comparable to the shot-noise level such that $n_{F,\text{corr}}=2$. Thus, even with higher bandwidth IR detectors, the \textbf{photonic correlator should have a limited impact on the performance of the overall system} (factor 2 for $100~\text{GHZ}$ detectors).

In addition, the photonic correlator shows a linear response over 5 orders of magnitude of RF input power, even in the presence of strong imbalance between its two RF inputs. This means that the RF gains do not need to be adjusted to accommodate to astronomical targets with different magnitudes, and the signals from each telescope do not need to be balanced.    

\begin{figure}
	\centering
	\includegraphics[width=0.7\linewidth]{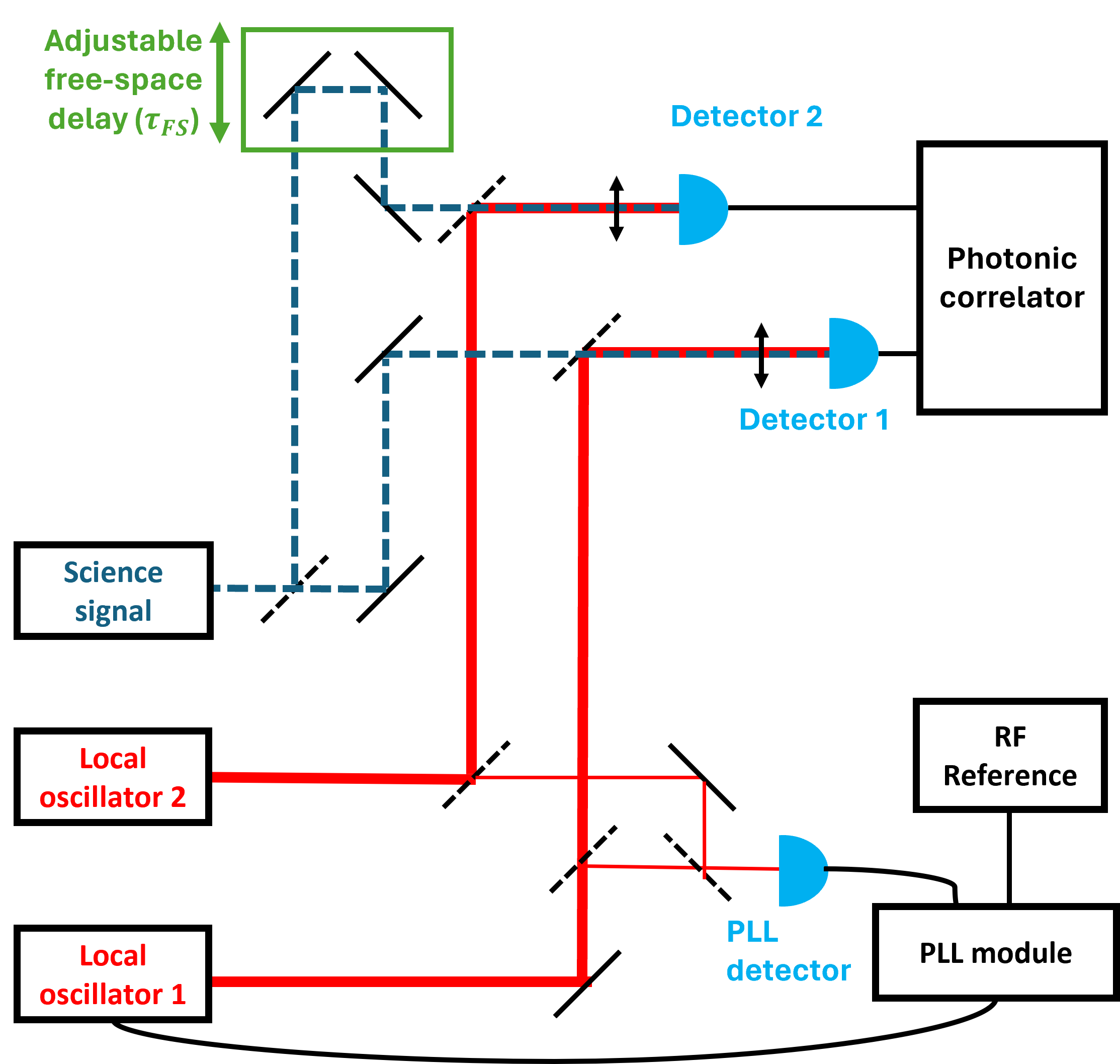}		
	\caption{Scheme for the implementation of separate LOs on each IR detector. A PLL is used to control the frequency and phase difference between the two LOs.}
	\label{fig:PLL}
\end{figure}

\subsection{Delay compensation}

we have shown that delay compensation between the telescopes could be achieved directly in the photonic correlator. In practice, assuming a $40~\mathrm{GHz}$ detection bandwidth, the coherence length of the signal would be on the order of one centimeter. Such precision should not be a problem for fibered delay lines at telecom wavelength. One of the possible limits for the delay compensation to be done with fibers is the differential dispersion. For single mode fibers with $17~\text{ps/km/nm}$ dispersion, considering signals with $40~\text{GHz}$, the delay can be compensated for, without significant impact from dispersion, up to $5~\text{km}$. Above this limit, the use of dispersion compensation fibers should be considered. Thus, \textbf{heterodyne interferometry could utilize kilometric fiber delay lines at telecom wavelengths instead of the mid-infrared free-space delay lines in direct interferometry}.

\subsection{Measurement of the interferometric phase}

Our current system does not recover the interferometric phase $\Delta\Phi_0$ of the signal and can only perform incoherent integration of the interferometric signal. From equation (\ref{eq:BPD_natural_frequency}), we see that $\Delta\Phi_0$ can be recovered from the correlator output signal as $\Delta\Phi_0=\frac{1}{2}\left( \Phi_{+}-\Phi_{-}\right)$ where $\Phi_{+}$ and $\Phi_{-}$ are the phases of the $f_-$ and $f_+$ frequencies.

In an on-sky system, $\Delta\Phi_0$ would carry the astronomical phase $\phi_\text{V}$ (\cite{Bourdarot2021}), the overall phase difference due to the path delay $\Delta\tau$ in the system, the atmosphere and the phase difference between the LOs at each telescope. Absolute measurement of the astronomical phase would not be possible, but phase closure information could be used in a system with more than two telescopes if phases in the system are stabilized.

In our current system, the phase difference between the LOs is not stabilized because of the dither mirror moving back and forth which creates the $f_{12}$ frequency shift. In ISI, the LOs were stabilized with Phase-Lock Loops (PLL). In the sort term, we plan to use a similar approach in our set-up with a separate LO for each detector and a PLL, as shown on Fig.~\ref{fig:PLL}. This would enable us to chose precisely the frequency difference $f_{12}$ between the two LOs and to maintain a stable phase difference between them based on an external phase reference. In the long term, we believe phase stabilization could be achieved for larger separations by synchronizing the local oscillators at the different telescopes. This could be done through a regular telecom fiber link at $1.5~\mathrm{\mu m}$, as demonstrated by \cite{argence2015} on a $10.3~\mathrm{\mu m}$ QCL which was stabilized down to $0.06~\mathrm{Hz}$ over $43~\mathrm{km}$. 

Regarding atmospheric turbulence and vibrations in the system, an on-sky system would need to operate with chunks of integration that are smaller than the coherence time of the atmospheric turbulence.
In the perspective of a long baseline on-sky heterodyne interferometer such as Planet Formation Imager, the use of an external fringe tracker, would enable coherent integration of the interferometric signal over much longer period, similar to what is performed with the GRAVITY instrument, thus drastically improving the SNR of the system and decreasing the detection limit.

The use of coherent integration would enable us to reduce the detection limit by a factor $11.5$ for one hour of integration time\footnote{Since one hour corresponds to $18000\times0.2~\text{s}$, the SNR is increased by a factor $\sqrt[4]{18000}\simeq 11.5$.}, down to $3~\text{Jy}$ for the QWIP detectors and $0.04~\text{Jy}$ for the ideal detectors at $100~\text{GHz}$ bandwidth. With perfect detectors, the detection limit would comparable to the current one hour  detection limit of the MATISSE instrument at the VLTI on one of its spectral bin (N band with $R=220$ resolution in GRA4MAT mode, which allows for coherent integration of the interferometric signal thanks to GRAVITY's fringe tracker) (\cite{lopez2022,petrov2024}).

\subsection{Wavelength multiplexing}

In our current system, we only detect the fraction of the incoming astronomical signal that falls inside the bandwidth of the detectors. Increasing the detector bandwidths reduces the detection limit of system up to approximately $100~\text{GHz}$, set by the current state of RF and telecom technologies. In comparison, the astronomical N band covers a frequency range of roughly $10~\text{THz}$. To collect more signal, a multiplexing technique with $N_\lambda$ spectral channels can be envisioned. In \cite{ireland2014}, the authors proposed to pave the N band with several thousand of detectors. Since each channel is independent, the total SNR of a multiplexed system scales as $\sqrt{N_\lambda}$. This shows that there is an exact SNR equivalence (at least in theory) between having $N_\lambda$ detectors with $\Delta\nu_{\mathrm{det}}$ bandwidth and having a single detector with $N_\lambda\Delta\nu_{\mathrm{det}}$ bandwidth.

Fifty channels would be needed to cover the N band with $\Delta\nu_{\mathrm{det}}=100~\text{GHz}$. In practice, going from a single spectral channel to $N_\lambda$ channels requires multiplying the number of detectors, the number of local oscillators, the number of amplifiers and the number of correlation operation by $N_\lambda$. For the local oscillator, the use of mid-infrared frequency combs could solve the issue because. A laser frequency combs is essentially equivalent to having a large number of phase-locked lasers with a given frequency spacing (\cite{schliesser2012a}). The astronomical signal and the comb would have to be mixed and they dispersed on a line of detectors. Since superimposing local oscillators of a single detector does not improve the SNR of the heterodyne detection, each comb line would have to fall on a different detector. The resulting heterodyne signals would need to be individually amplified. For the photonic correlation, we would not necessarily need to use $N_\lambda$ fibers because we can utilize the widely used wavelength-division multiplexing techniques. First, each spectral channels would be encoded onto a different telecom wavelength. Then, the channels would be multiplexed in a single fiber for transport. Finally, the channels would be demultiplexed and correlated. However we would still need $N_\lambda$ times more intensity modulators, AOFSs and balanced photodiodes.

It should be noted that performing the correlation operation on a single broadband balanced detector without demultiplexing the channels is possible with well-chosen acousto-optic frequency shifts. However, this would reduce the SNR proportionally to the number of channels because each correlation operation would add broadband noise ($w_1w_2$ terms) affecting all correlation peaks.

\section{Conclusions}

We have demonstrated the capacity to measure the coherent flux of a
$10.6~\mathrm{\mu m}$ low-coherence source using a two telescope heterodyne
interferometry approach with photonic correlation. The photonic correlator,
solely built from off-the-shelf telecom components, is able to correlate
signals with $10~\mathrm{GHz}$ bandwidths, directly extendable to
$40~\mathrm{GHz}$.

We characterized the response of the photonic correlator to coherent signals and its robustness to noise. The current overall detection limit of our system is $140\pm 30~\mathrm{fW}$ for an integration time of $0.2~\mathrm{s}$, limited by our IR detectors. We showed that free-space delay between the two IR detector could be compensated by fiber delay inside the correlator. This represents a significant demonstration of the feasibility of an all-fibered mid-infrared interferometer.

For a total incoherent integration time of $3600~\text{s}$, the astronomical detection limit of our system is $130~\text{Jy}$ for an $8~\text{m}$ telescope. The use of state-of-the-art QWIP detectors could enable us to decrease this limit down to approximately $35~\text{Jy}$ and down to $3~\text{Jy}$ with coherent integration. Coherent integration would also bring the possibility to measure the astronomical phase or phase-closure from the output of the photonic correlator. In addition, spectral multiplexing using laser frequency combs as local oscillator would further improve considerably the sensitivity (\cite{Bourdarot2021}).

\bigskip
\emph{We acknowledge financial support from \emph{ENS-THALES Chair}, from \emph{LabEx FOCUS ANR-11-LABX-0013}, from \emph{ASHRA} and from \emph{INSU CSAA}.
      The authors gratefully acknowledge the support from Sylvain Rochat, Jérémy Ceszkowski, Alain Delboulbé and David Gillier for setting up the experiment and Bruno Maillard and Marie-Hélène Sztefek for handling the financial and administrative part.}

\bibliographystyle{plain}
\bibliography{biblio}

\clearpage
\glsaddall
\printnoidxglossaries

%\newpage
\begin{appendix}
\section{QCLs}
\label{app:QCL}

\begin{figure*}
	\centering
	\begin{minipage}{0.49\hsize}
		\includegraphics[width=1\linewidth]{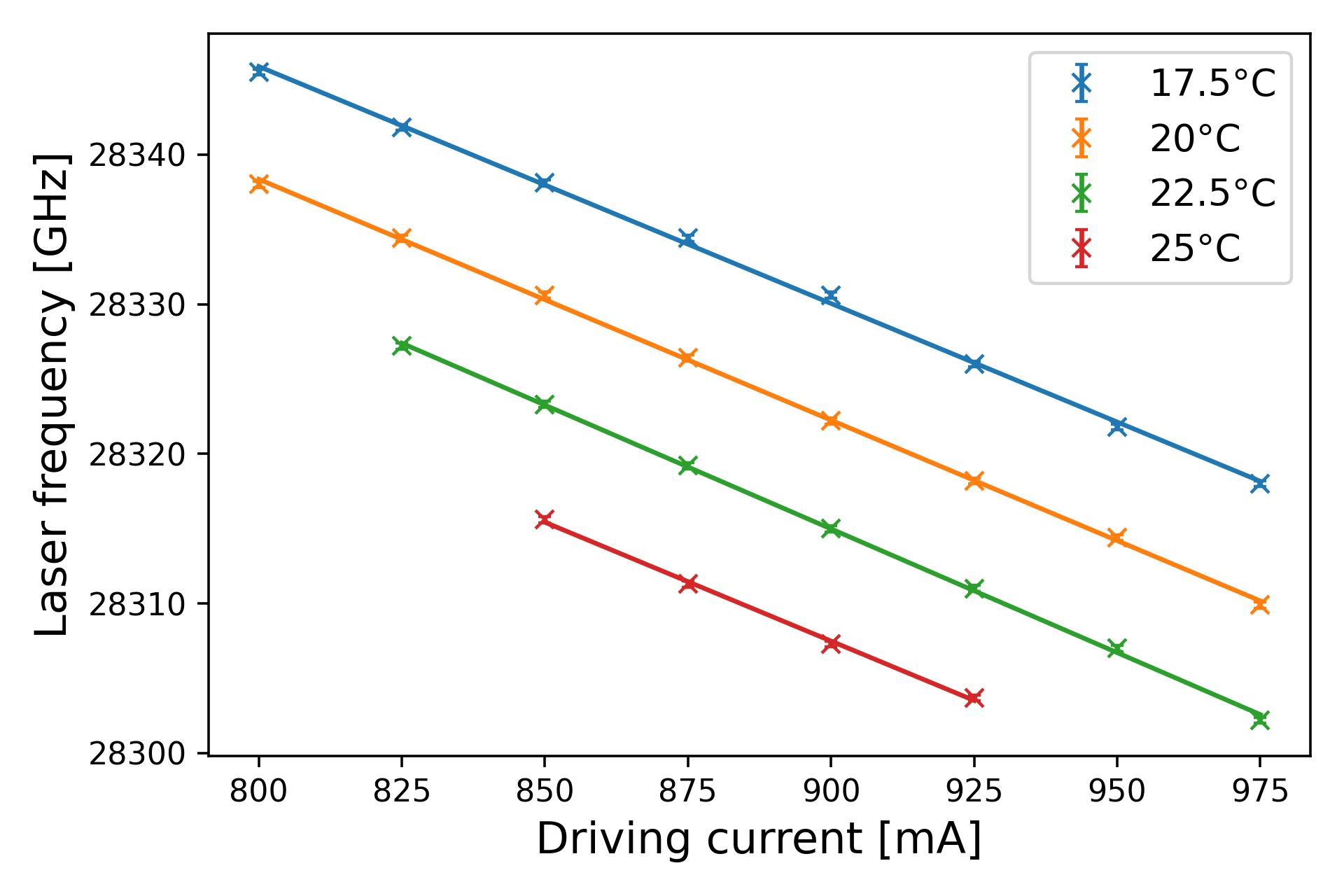}
	\end{minipage}
	\hfill
	\begin{minipage}{0.49\textwidth}
		\includegraphics[width=1\linewidth]{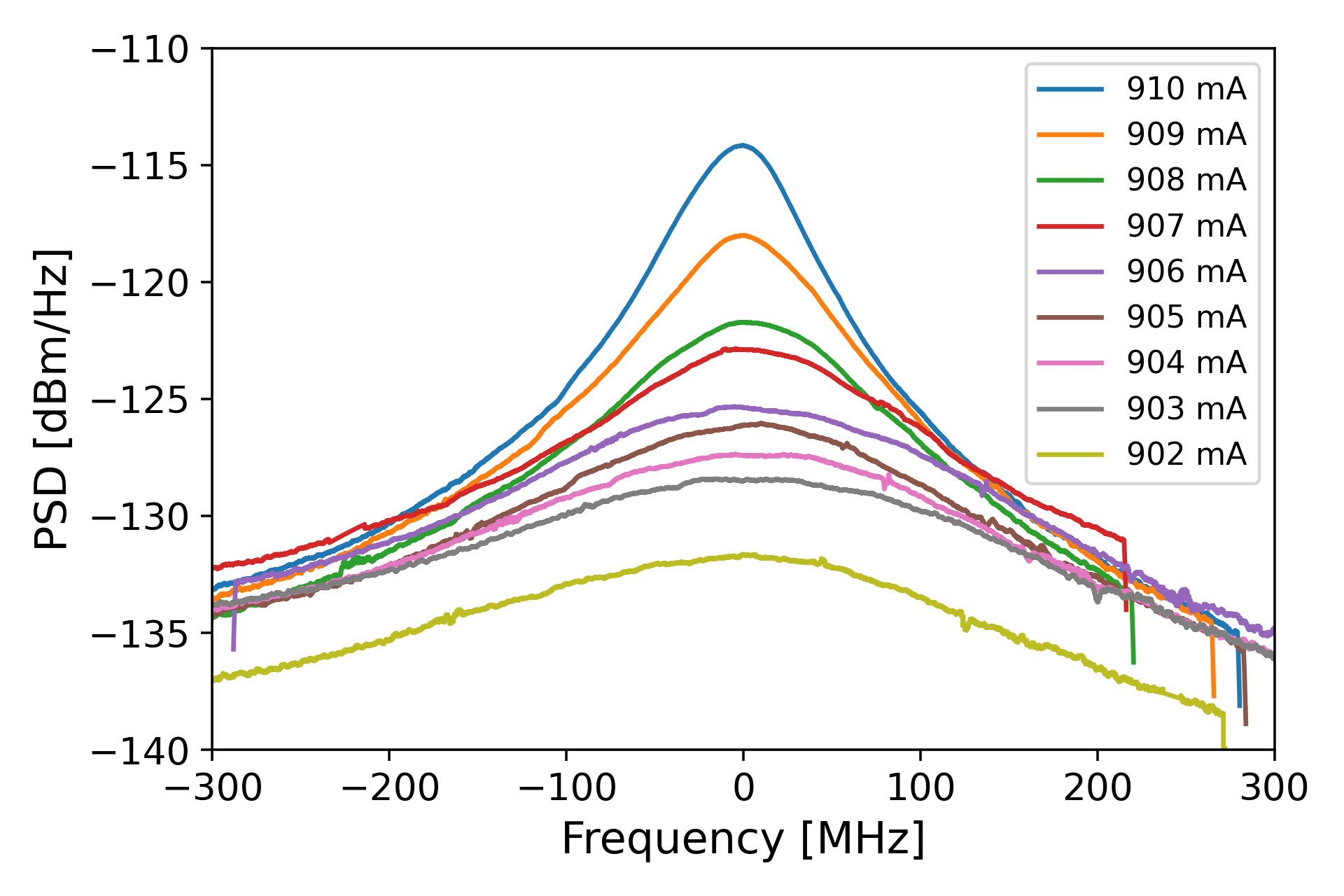}
	\end{minipage}
	\caption{\emph{(Left)} QCL B frequency as a function of driving current for different temperatures. \emph{(Right)} PSD of the ASE signal of QCL A for different driving current, measured via heterodyne detection. All the spectra were centered at $0$.} 
	\label{fig:QCL}
\end{figure*}

We use Quantum Cascade Lasers (QCLs) as LO and science sources in our set-up. The frequency of a QCL can be adjusted by changing its driving current. In Fig.~\ref{fig:QCL} \emph{(Left)}, the frequency of QCL B was measured as a function of driving current. In free-running laser mode, QCLs have narrow optical bandwidths on the order of $\sim 1~\text{MHz}$. When precisely controlled, even at a distance of a few tens of kilometers, QCLs can reach sub-Hz frequency precisions, as demonstrated by \cite{argence2015}, making them ideal LOs for heterodyne interferometry applications.

\subsection{Amplified Spontaneous Emission regime}

The Amplified Spontaneous Emission (ASE) regime of a laser corresponds to the emission of radiation below the laser threshold. The ASE regime exhibits a much weaker optical power than laser mode (on the order of $\sim 1$ to $\sim 10~\mathrm{nW}$) with a much wider optical bandwidth (on the order of $\sim 10$ to $\sim 100~\mathrm{MHz}$). The ASE mode can be used as an intermediate between a fully coherent laser and a fully incoherent black-body source for the science source. The main advantage of using a QCL in ASE regime is the possibility to align and calibrate the system in laser mode before switching the ASE without disturbing the alignment.  

In Fig.~\ref{fig:QCL} \emph{(Right)}, we show the spectral shape of QCL A in ASE regime for different driving frequencies. The measurements were obtained by averaging power spectra from recorded heterodyne beating series between QCL B in laser mode and the QCL A in ASE regime. We can chose the driving current to obtain different optical powers or coherent lengths for the science source. For example, with a driving current of $902.0~\text{mA}$, we measured a bandwidth of $300~\text{MHz}$ which would correspond to a coherence length of $32~cm$ assuming Lorentzian shape.
%%%%%%%%%%%%%
%%%%%%%%%%%%%
%%%%%%%%%%%%%

\section{Observable extraction from correlator output}

\label{app:correlator_signal_extraction}

In an ideal set-up, our main observable $\mathcal{H}$ is given by equation (\ref{eq:P_corr}) as the sum of the $f_-$ and $f_+$ components of the BPD signal $i_{\text{BPD}}$ where the correlation product is encoded. However, in our current set-up, the natural fringe frequency $f_{12}=70~\mathrm{Hz}$ is generated by a dither mirror moving back and forth. The vibrations, accelerations, decelerations and pauses in the movement of the mirror disperse the correlation information over a wider range of frequencies (typically $[174.5,175.5]~\mathrm{kHz}$, as shown in Fig.~\ref{fig:correlation_peak}.

We extract our observable $\mathcal{H}$ using equation (\ref{eq:P_corr_extraction}) where $\text{PSD}_{\text{background}}$ is the correlator output PSD in the absence of the science signal.

\begin{equation}
\begin{aligned}
    \mathcal{H}=&\int_{174.5~\mathrm{kHz}}^{175.5~\mathrm{kHz}}(\text{PSD}_{\text{corr}}(\nu)-\text{PSD}_{\text{background}}(\nu))\mathrm{d}\nu
    \\
    &-\frac{3}{4}(\text{PSD}_{\text{corr}}(175~\mathrm{kHz})-\text{PSD}_{\text{background}}(175~\mathrm{kHz}))
\end{aligned}
\label{eq:P_corr_extraction}
\end{equation}

The $175.000~\text{kHz}$ contribution to the correlation power must be divided by 4 to account for the fact that other frequency contributions at $175~\text{kHz}+\Delta f$ are split between $175~\text{kHz}-\Delta f$ and $175~\text{kHz}+\Delta f$ (factor 2 in amplitude, leading to a factor 4 in power).

\subsection{$\mathcal{H}$ extraction for low power}

In the detection limit measurement, the central peak from the background is typically $10^4$ greater than the signal peak as shown by Fig.~\ref{fig:correlator_output}. Therefore, if the correlation power to be detected is too low, the noise contributions from $\text{PSD}_{\text{background}}$ around the $175~\mathrm{kHz}$ peak will significantly degrade the results. 

We solved this issue only summing the frequency components in the $[174.925,174.930]~\mathrm{kHz}$ and $[175.070,175.075]~\mathrm{kHz}$ ranges were most of the correlation signal is located, with good SNR. The result is then corrected by a factor $0.61\pm 0.03$ which is the fraction of the correlation power contained in the limited frequency ranges compared with the total $[174.5,175.5]~\mathrm{kHz}$ range. This factor was measured in the presence of a high correlation power. With that method, $\mathcal{H}$ is given by equation (\ref{eq:P_corr_extraction_low_power}).

\begin{equation}
\begin{aligned}
    \mathcal{H}=&\frac{1}{0.61}\left(\int_{174.925~\mathrm{kHz}}^{174.930~\mathrm{kHz}}(\text{PSD}_{\text{corr}}(\nu)-\text{PSD}_{\text{background}}(\nu))\mathrm{d}\nu\right.
    \\
    &+\left.\int_{175.070~\mathrm{kHz}}^{175.075~\mathrm{kHz}}(\text{PSD}_{\text{corr}}(\nu)-\text{PSD}_{\text{background}}(\nu))\mathrm{d}\nu\right)
\end{aligned}
\label{eq:P_corr_extraction_low_power}
\end{equation}

\section{System calibration}
\label{app:correlator_laser_callibration}
\begin{figure}[]
	\centering
    \includegraphics[width=0.7\linewidth]{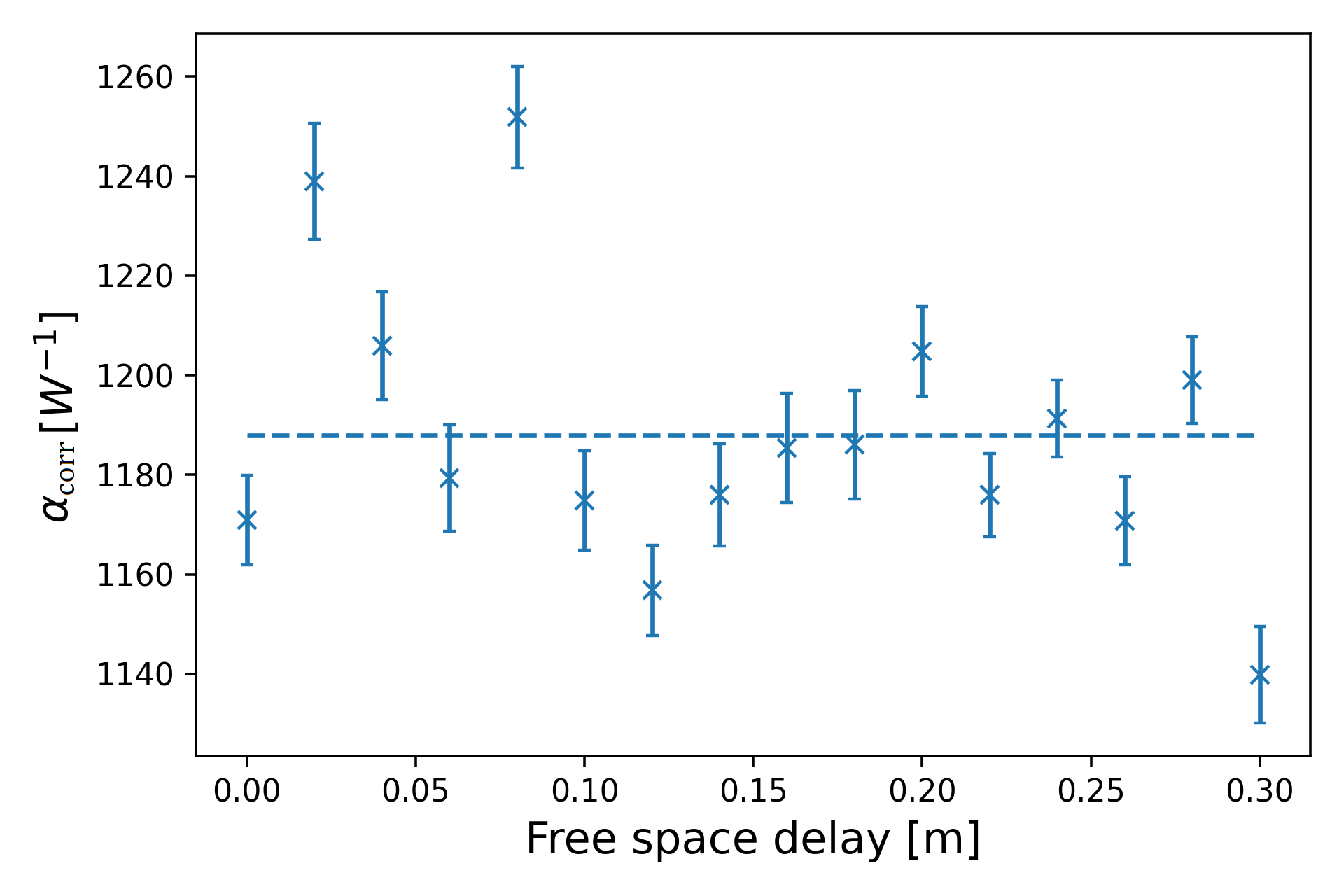}
	\caption{Measurement of parameter $\alpha_{\text{corr}}$ at different free-space delay $\Delta\tau_{\text{FS}}$ positions.}
	\label{fig:laser_calibration}
\end{figure}

One limitation of the correlator response measurement with signals generators comes from the limited frequency range. Our signal generator could not generate signals above $120~\text{MHz}$, but we operate the system using intermediate frequencies mostly between $200~\text{MHz}$ and $500~\text{MHz}$.

We calibrated the response of the correlator at $300~\text{MHz}$ by measuring the correlation of laser-laser heterodyne detection with different free-space delays $\Delta\tau_{\text{FS}}$. The goal is to measure the value of $\alpha$.

The two QCLs were set to laser mode with an intermediate frequency of $300~\mathrm{MHz}$. The driving current of the LO QCL was set to $900.1~\text{mA}$. The driving current of the science signal QCL was set to $915.0~\text{mA}$.

We used 50/50 RF power splitters to measure the RF powers $P_{\text{RF},1}$ and $P_{\text{RF},2}$ at the output of the IR detectors while sending the same amount of RF power to the photonic correlator to measure $\mathcal{H}$. Knowing we have $\gamma_{12}=1$ for a laser source, we can recover the value of $\alpha_{\text{corr}}$ from equation (\ref{eq:degree_of_coherence}). We performed the measurement at different delays $\Delta\tau_{\text{FS}}$, as shown on Fig.~\ref{fig:laser_calibration}. We obtained values in the $[1140,1260]~\mathrm{W}^{-1}$ range, with an average value of $1190\pm 10~\mathrm{W}^{-1}$. As expected, the value of $\alpha_{\text{corr}}$ does not seem to be correlated with the free space delay. We believe the dispersion of the values outside their relative uncertainty comes from the wander of the frequency difference between the QCLs which are not phase-locked. Therefore, we take some margin and consider $\alpha_{\text{corr}}=1190\pm 70~\mathrm{W}^{-1}$.

%%%%%%%%%%%%%
%%%%%%%%%%%%%
%%%%%%%%%%%%%
\section{Numerical simulation of the photonic correlator}
\label{app:simulation}

\begin{figure*}
	\centering
	\includegraphics[width=1\linewidth]{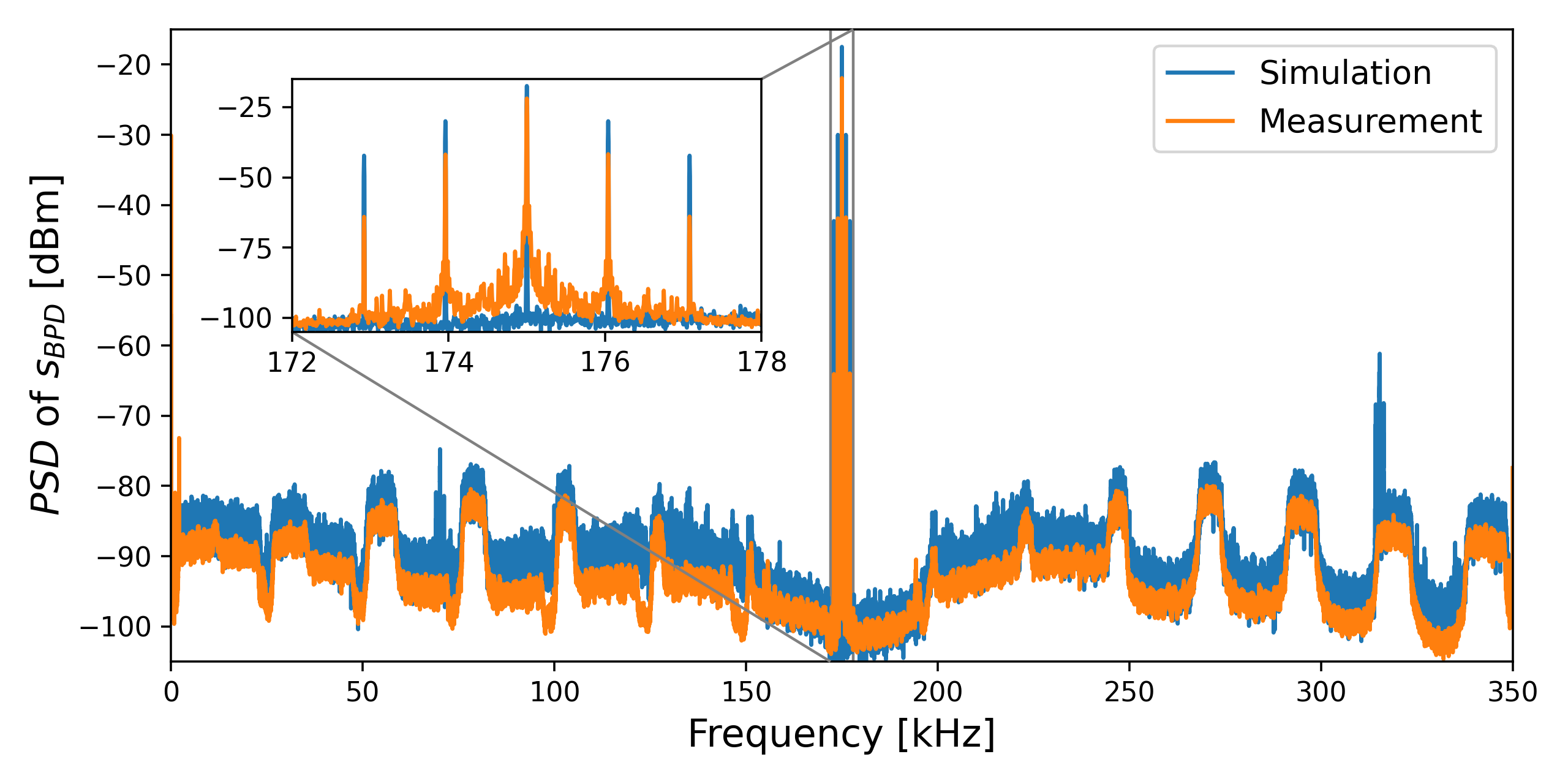}
	\caption{Comparison between the measured power spectrum of $i_{\text{BPD}}$ in the absence of input signals and the simulation.} 
	\label{fig:noise_floor}
\end{figure*}

\begin{figure}[]
	\centering
    \includegraphics[width=0.7\linewidth]{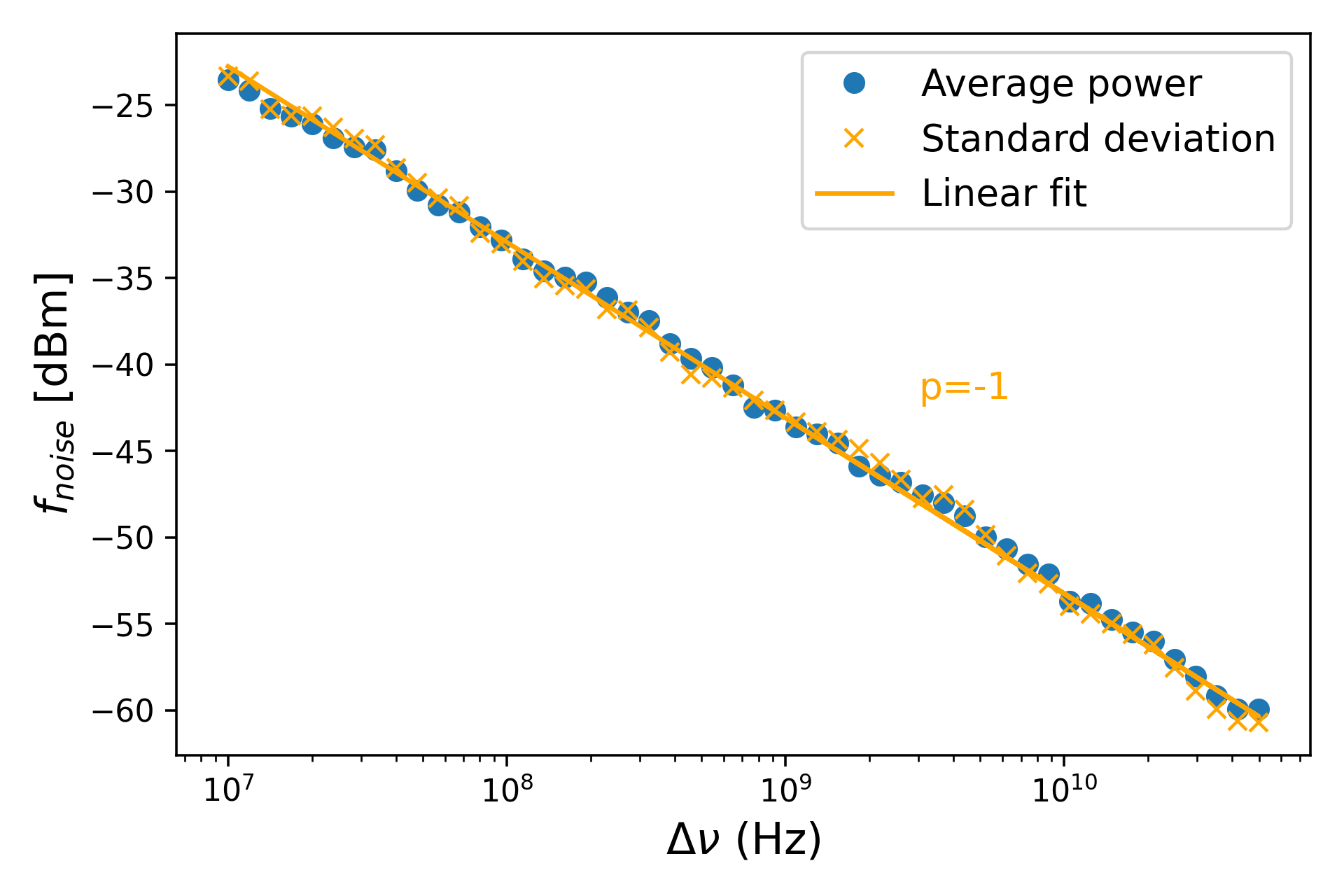}
	\caption{$f_{\text{noise}}$ as a function of RF noise bandwidth $\Delta\nu$. The RF power was kept constant at $0~\text{dBm}=10^{-3}~\text{W}$ for all $\Delta\nu$ values.}
	\label{fig:noise_bandwidth}
\end{figure}

We used a numerical model which simulates the main equations of the photonic correlator (notably equations (\ref{eq:MZM},\ref{eq:AOFS},\ref{eq:I_+-},\ref{eq:BPD})) and compared the results with the experimental data from the bench. The simulation integrates the physical parameters of the bench such as the transmission coefficient of the fibers and components, $V_{\pi,k}$ values of the Mach-Zehnder intensity modulator, responsivity of the balanced photodetector, etc. It also includes typical noises from the components of the correlator.

Coherent or incoherent wide-band RF signals can be randomly generated and used as RF inputs. In the case of coherent signals, a frequency shift between $h_1$ and $h_2$ is applied to simulate the presence of the natural fringe frequency $f_{12}$.

We used the simulation to confirm our interpretation of the $f_{\text{noise}}$ function and study the impact of the different elements of the correlator on the overall system. The simulation shows very good agreement with the measurement. As shown by Fig. \ref{fig:noise_floor}, it can notably reproduce the spectral features of the $i_{\text{BPD}}$ signal. %The simulation shows a good agreement with the experimental data, as shown on Fig.~\ref{fig:correlation_nep}~\emph{(Right)}.

\subsection{Noise bandwidth}
\label{app:simulation_noise_bandwdith}

We used to simulation to study the variation of $f_{\text{noise}}$ with the noise bandwidth $\Delta\nu_{\text{noise}}$. We expect $f_{\text{noise}}$ to be inversely proportional to $\Delta\nu_{\text{noise}}$ because the product of two white noises $w_1$ and $w_2$ yield a triangular noise whose spectra decreases linearly and reaches $0$ at $2\Delta\nu_{\text{noise}}$ (see
\cite{Kish2012}). %Therefore, for a given RF power, the noise contribution of the $w_1w_2$ product at the fringe peak frequency $f_{12}$ will be lower if $\Delta\nu_{\text{noise}}$ is higher because the power will be distributed among a wider range of frequencies.

We verified the validity of this hypothesis with the numerical simulation of the photonic correlator. We generated $50$ sequences of $20~\mathrm{\mu s}$ of white noise with simulation bandwidth of $100~\text{GHz}$. We filtered the noise inputs to obtain different noise bandwidths $\Delta\nu_{\text{noise}}$ from $10~\text{MHz}$ to $50~\text{GHz}$. The RF power of the noises were normalized to $0~\text{dBm}=1~\text{mW}$ after filtering. The average value of $P_{\text{corr}}$ and the noise level (the standard deviation of $P_{\text{corr}}$) are shown on Fig.~\ref{fig:noise_bandwidth} as a function of $\Delta\nu_{\text{noise}}$. We obtain a $p=-1$ slope in logarithmic scale corresponding to the expected inverse proportionality between the noise level and $\Delta\nu_{\text{noise}}$. This inverse proportionality can only be applied to the noise that originates from $w_1$ and $w_2$ and cannot be applied to the
noise floor of the correlator. Thus, we can write $f_{\text{noise}}$ in the form of equation (\ref{eq:f_noise}) which is reminded here as equation (\ref{eq:f_noise_appendix}).

\begin{equation}
    f_{\text{noise}} \simeq \alpha_{\text{IR}}\left(A+B\times P_{\text{RF,noise}}+\frac{C}{\Delta\nu_{\text{noise}}}\times{P_{\text{RF,noise}}}^2\right)
	\label{eq:f_noise_appendix}
\end{equation}

%%%%%%%%%%%%%
%%%%%%%%%%%%%
%%%%%%%%%%%%%
\section{Coherence envelop measurement}
\label{app:coherence_envelop}
We measured the coherence envelop of QCL A in ASE mode by measuring the mutual coherence $\Gamma_{12}$ at different free space delays $\Delta\tau_{\text{FS}}$. We set QCL A and QCL B driving currents respectively to $902.0~\text{mA}$ and to $890.0~\text{mA}$.

Calculating the degree of coherence $\gamma_{12}$ from $\Gamma_{12}$ requires measuring $P_{\text{RF},1}$ and $P_{\text{RF},2}$, as stated by equation (\ref{eq:degree_of_coherence}). Since changing the delay $\Delta\tau_{\text{FS}}$ can modify the mode coupling $\mathcal{C}_2$ between the LO and the science source on detector 2, measurement of $P_{\text{RF},2}$ is necessary every time the delay is changed to obtain calibrated results.

Because of the low optical flux of the ASE which was used as wide-band science source, we resorted to using a calibration laser. Instead of measuring of $P_{\text{RF},1}$ and $P_{\text{RF},2}$ every time the delay was changed, we measured the RF power of a laser-laser heterodyne intermediate frequency $P_{\text{RF},\text{cal}}$ and scaled the results accordingly.

The $\gamma_{12}$ values were fitted with a Lorentzian function and normalized to 1. We chose to set QCL A driving current to a specific value of $917.0~\text{mA}$ for all the laser-laser calibration measurements. The current of QCL B was adjusted to obtain a reference heterodyne beating at $300~\text{MHz}$. In our case, due to the instability of the laser frequencies, the value of the driving current for the QCL A varied between $901.6$ and $902.4~\text{mA}$ to retrieve the $300~\text{MHz}$ beating\footnote{Such variation of the laser driving current can cause a variation of the local oscillator intensity by less than $1\%$ which is negligible compared with our sources of uncertainty that will be discussed later. This $1\%$ is lower than the variation of the response of the detector if QCL B current was set to a precise value.}.

The overall measurement protocol can be summarized as follows:
\begin{enumerate}
    \item Move detector 2 at desired position and adjust the alignment of the QCLs if needed, without modifying the alignment for detector 1.
    \item Add optical density 3 in front of the local oscillator to reduce its power by a factor $10^3$ and avoid saturation. The optical density was mounted on a flip-mount for repeatability.
    \item Set the QCLs to specified calibration mode.
    \item Measure $P_{\text{RF},2}$. The typical uncertainty we obtained on such a measurement was $2$ to $5\%$.
    \item Set the QCLs to ASE-laser mode.
    \item Remove optical densities.
    \item Measure the correlation signal $P_{\text{corr}}$ at the correlator output. The typical uncertainty we obtained on such a measurement was $2$ to $10\%$.
    \item Divide the $\Gamma_{12}$ value by the laser reference $P_{\text{RF},2}$.
\end{enumerate} 

In the end, after fitting the results and normalizing them to one, we obtain Fig.~\ref{fig:coherence_envelop}.

\section{Detection limit measurement}
\label{app:detection_limit}

We set the set-up so that one QCL (the LO) is in laser mode and the second one in ASE mode (the science signal). The resulting optical power of the ASE is $6.9\pm1.5~\text{nW}$.

We were able to retrieve some correlator signal with total optical densities of $3.5$ ($3.2\times 10^{-4}$ transmission). We extracted the correlation power in the $[174.925,174.930]~\text{kHz}$ and $[175.070,175.075]~\text{kHz}$ ranges were the SNR is maximal, knowing it carries $61\pm3\%$ of the total correlation power. We obtained an average value of $P_{\text{corr}}=(4.4\pm0.2)\times 10^{-9}~\text{W}$ which, taking the $61\pm3\%$ factor into account, corresponds to $P_{\text{corr}}=(7.3\pm0.3)\times 10^{-9}~\text{W}$. The standard deviation of the correlation power among the 250 measurements is $5.8\times 10^{-9}~\text{W}$.

Using equations (\ref{eq:correlator_calibration}) and (\ref{eq:f_corr}), we retrieved the coherent flux that correspond to the level of correlator output. We obtained $|F_c|=140\pm 30~fW$. 

Assuming $\Gamma_{12}=1$ and taking into account the optical densities that were added, we can calculate the optical power $F_c$ of the ASE source. We obtain $F_c=5.3\pm1.0~\text{nW}$ which is compatible with the expected $F_c=6.9\pm1.5~\text{nW}$.

\end{appendix}

\end{document}